\documentclass{aa}
\usepackage{amsmath}
\usepackage{amssymb}
\usepackage{graphicx}
\usepackage{txfonts}
\usepackage{multirow}
\usepackage{float}
\usepackage{orcidlink}
\usepackage{xcolor}
\hypersetup{colorlinks = true, allcolors = blue}
\usepackage{CJKutf8}
\usepackage{booktabs}
\usepackage{hyperref}
\usepackage{ragged2e}
\usepackage{bm}
\usepackage{placeins}
\usepackage{capt-of}

\begin{document}
   \title{Search for quasar pairs with \textit{Gaia} astrometric data.}
   \subtitle{II. Photometric redshift prediction with machine learning for the MGQPC catalogue}
   \titlerunning{Search for Quasar Pairs with \textit{Gaia} Astrometric Data. II.}
   \author{
        Xingyu Zhu (\begin{CJK}{UTF8}{gbsn}{朱星宇}\end{CJK}) \inst{\ref{BNU_PA},\ref{BNU_FIAA}} \orcidlink{0009-0008-9072-4024}
        \and
        Qihang Chen (\begin{CJK}{UTF8}{gbsn}{陈启航}\end{CJK}) \inst{\ref{BNU_PA},\ref{BNU_FIAA}} \orcidlink{0009-0006-9345-9639}
        \and
        Liang Jing (\begin{CJK}{UTF8}{gbsn}{荆亮}\end{CJK}) \inst{\ref{BNU_PA},\ref{BNU_FIAA}} \orcidlink{0000-0003-1188-9573}
        \and
        Zhuojun Deng (\begin{CJK}{UTF8}{gbsn}{邓卓君}\end{CJK}) \inst{\ref{BNU_PA},\ref{BNU_FIAA}} \orcidlink{0009-0008-8080-3124}
        \and
        Jun-Qing Xia (\begin{CJK}{UTF8}{gbsn}{夏俊卿}\end{CJK})\inst{\ref{BNU_PA},\ref{BNU_FIAA}}
        \and
        Yanxia Zhang (\begin{CJK}{UTF8}{gbsn}{张彦霞}\end{CJK}) \inst{\ref{NAOC}} \thanks{Corresponding author: \href{mailto:zyx@bao.ac.cn}{zyx@bao.ac.cn}} \orcidlink{0000-0002-6610-5265}
        \and
        Jianghua Wu (\begin{CJK}{UTF8}{gbsn}{吴江华}\end{CJK}) \inst{\ref{BNU_PA},\ref{BNU_FIAA}} \thanks{Corresponding author: \href{mailto:jhwu@bnu.edu.cn}{jhwu@bnu.edu.cn}} \orcidlink{0000-0002-8709-6759}
        }
   \institute{
              School of Physics and Astronomy, Beijing Normal University, Beijing 100875, PR China\\
              \email{jhwu@bnu.edu.cn}
              \label{BNU_PA}
              \and
              Institute for Frontier in Astronomy and Astrophysics, Beijing Normal University, Beijing 102206, PR China
              \label{BNU_FIAA}
              \and
              National Astronomical Observatories, Chinese Academy of Sciences, Beijing 100101, PR China
              \label{NAOC}
            }
   \date{Received XXXX, XXXX; accepted XXXX, XXXX}

    \abstract
    {The identification of physically associated kiloparsec-scale quasar pairs is important for understanding galaxy evolution, the growth of supermassive black holes, and co-evolution of supermassive black holes and their host galaxies. However, their intrinsic rarity and high contamination from stellar superpositions and projected alignments in photometric catalogues necessitate efficient selection methods. This challenge is amplified as wide-field imaging surveys such as the Legacy Survey of Space and Time (LSST) and \textit{Euclid} will discover vast numbers of quasar candidates, most lacking immediate spectroscopic follow-up, making accurate photometric redshifts (photo-$z$) essential for pre-selection.}
   {This work aims to develop a robust machine learning framework to produce accurate photo-$z$ point estimates and well-calibrated probability density functions (PDFs) for quasars. The primary application is to systematically identify high-probability quasar pair candidates within the MGQPC (quasar pair candidates derived from the cross-match of MQC and Gaia) catalogue by assessing the redshift consistency of pair components.}
   {We constructed two large, spectroscopically confirmed quasar samples with multi-wavelength photometry: an SDSS-based sample (KSTS) and a DESI-LS-based sample (KSTD). We employed the \textsc{CatBoost} algorithm for point-estimate photo-$z$ regression and the \textsc{FlexZBoost} framework for full photo-$z$ PDF estimation. A feature selection process guided by SHAP (SHapley Additive exPlanations) was implemented to identify the most predictive photometric features, including colours and magnitudes from optical (e.g. $u, g, r, i, z, G, G_{\rm BP}, G_{\rm RP}$) and infrared ($W1, W2$) bands.}
   {Our photo-$z$ workflow achieves a robust performance, with a normalised median absolute deviation of $\sigma_{\mathrm{NMAD}}=0.036$ and an outlier fraction of 5.6\% on the test sample. The resulting photo-$z$ PDFs are well calibrated, as verified by probability integral transform diagnostics. Control tests indicate that the better performance of the SDSS-based sample is primarily attributable to the inclusion of the $u$ band, although survey-dependent photometric differences in LS10, including bright-source systematics, may also play a contributing role. Applying this trained model to the MGQPC catalogue, we identified 185 high-probability quasar pair candidates based on the consistency between their photo-$z$ estimates. This candidate list included 20 systems subsequently confirmed as genuine physical pairs by independent spectroscopic observations.}
   {We demonstrate that modern machine learning techniques, combining gradient-boosted trees for point estimates and conditional density estimation for PDFs, can deliver precise and reliable photo-$z$s for quasars. This enables the efficient filtering and prioritisation of rare quasar pairs from large photometric catalogues. The resulting photo-$z$ catalogue for the MGQPC provides a valuable resource for future spectroscopic follow-up campaigns to study dual supermassive black holes and their co-evolution. }      


    \maketitle

\section{Introduction} \label{sec1}
Quasar pairs or dual quasars (hereafter quasar pairs collectively) are those systems of two quasars with a projected transverse separation of $30~\mathrm{pc}\lesssim r_{\rm p}\lesssim 100~\mathrm{kpc}$ and radial velocity offset of $\left\lvert\Delta v\right\rvert \lesssim 2000~\mathrm{km\,s^{-1}}$ \citep[e.g. ][]{DeRosa2019QPreview,Pfeifle2025BigMAC}.  Such systems are a natural expectation of hierarchical galaxy assembly and provide laboratories for studying galaxy interactions and mergers, the co-evolution of supermassive black holes (SMBHs) and their hosts, as well as quasar environments and clustering \citep[e.g. ][]{Begelman1980BAGN,Impey1998QPLSS,YuQJ2002MNRASBSMBHevolution,Hennawi2006BQinSDSSclustering,Myers2008BQclustering,LiuX2011AGNpairfraction,ShenY2010BQclustering,ShenY2023QPfraction,Sandrinelli2014QPenvironment,Sandrinelli2018QPenvironment,HouMC2020LiuXAGNpairXstats}. They are closely connected to the formation and evolution of SMBH binaries and the `final-parsec' problem \citep[e.g. ][]{Merritt2013AGNevolution,WangJM2023BSMBHfinalparsec,AnT2022BSMBHVLBI,XuWC2024DAGNVLBA}. As plausible progenitors of SMBH binaries, they may contribute to the nanohertz (nHz) gravitational-wave background \citep{Kelley2019aQPnHzGW,ChenYF2020BSMBHevolutionGW,ShenY2023DSMBHnHzGW}. 

Despite their importance, quasar pairs remain observationally scarce, with only about 200 such systems that have been spectroscopically confirmed to date \citep[e.g. ][]{Pfeifle2025BigMAC}. Identifying additional kiloparsec-scale quasar pairs is challenging even within our adopted projected-separation limit. Close companions are limited by the angular resolution of typical ground-based imaging and are further affected by target-allocation incompleteness (e.g. fibre collisions) in modern multi-object spectroscopic surveys. Candidate lists may also be heavily contaminated by chance alignments, in particular by foreground-star superpositions \citep[e.g. ][]{LiuX2011AGNpairfraction,ShenY2023QPfraction}. Using the SDSS DR16Q quasars \citep{Lyke2020} together with Gaia DR3 data, Jiang et al. (2025) showed that foreground-star superpositions dominate ($\gtrsim 80\%$) among Gaia-resolved close companions, and that the resolving-completeness-corrected double-quasar fraction is only $(5.7 \pm 0.3)\times10^{-4}$. These considerations highlight both the intrinsic rarity of genuine kiloparsec-scale quasar pairs and the need for high-purity candidate selection followed by efficient validation and follow-up prioritisation.

Motivated by the scarcity of quasar pairs and the challenges in searching for these systems, we initiated a \textit{Gaia} astrometry search for quasar pair candidates and spectroscopic follow-ups. In Paper~I, \citet{Chen2025GaiaPairsI} searched \textit{Gaia} sources with near-zero proper motions and near-zero parallaxes within a projected distance of 100 kpc to quasars and quasar candidates listed in the Million Quasar Catalogue (MQC), yielding 4\,112 candidate quasar pairs (denoted as MGQPCs). Here, the 100 kpc limit corresponds to a redshift-dependent angular-separation threshold rather than to a single fixed angular separation (see Fig.~9). The MQC served only as the parent catalogue for the Gaia-based candidate search in Paper~I, rather than as the redshift-truth source for the machine-learning training in the present work. Each MGQPC consists of two members: an MQC-listed quasar or quasar candidate (MGQPC\_A) and a nearby quasar candidate (MGQPC\_B). A recent statistical work \citep{Deng2025GaiaPairsIII} has identified 160 MGQPCs by cross-matching them with the spectroscopic data of the Dark Energy Spectroscopic Instrument Data Release 1 \citep[DESI DR1;][]{DESIdr1}. However, most of them are projected quasars. These new quasars are therefore close only in projection, with spectroscopic redshifts inconsistent with the physical-pair criterion adopted in this work (see Sect.~\ref{sec5.4}). To increase the success rate of our quasar pair confirmation, in this work, we tried to estimate the photometric redshifts (photo-$z$s) of the MGQPC\_Bs and to isolate those candidates with photo-$z$s close to the spectroscopic redshifts (spec-$z$s) of their companions MGQPC\_As.

Photo-z estimation is usually implemented via spectral energy distribution (SED) template fitting or machine learning (ML) \citep{Salvato2019}. The template-fitting methods compare multi-band photometry with libraries of SED models \citep[e.g. ][]{Arnouts1999,Benitez2000,Bolzonella2000,Brammer2008}, while the ML methods learn the mapping from photometric observables to redshifts from training data and can achieve a strong performance when large, representative spectroscopic samples are available \citep[e.g. ][]{Schmidt2020,Fu2024CatNorth,Roster2024PICZL}. 
However, ML models may generalise poorly outside the training domain and are sensitive to training-set representativeness \citep{Hastie2001}.

In this paper we present a dedicated ML photo-$z$ workflow for \textit{Gaia}-selected quasar pair candidates, with a particular focus on the MGQPC catalogue. For quasar pair applications, photo-$z$ enables uniform catalogue-level screening and follow-up prioritisation. Uncertainty quantification is crucial because redshift consistency between the two members depends not only on point estimates but also on the overlap of their redshift probability density functions (PDFs). We used the \textsc{CatBoost} gradient-boosted decision-tree regressor \citep{Prokhorenkova2018} to obtain point-estimate photo-$z$ values and a conditional density estimation (CDE) approach based on \textsc{FlexZBoost}/FlexCode \citep{IzbickiLee2017,Dalmasso2020,Schmidt2020} to infer full redshift PDFs. The models were trained on homogeneous spectroscopic quasar samples from SDSS \citep{York2000}, the Large Sky Area Multi-Object Fibre Spectroscopic Telescope \citep[LAMOST; ][]{cui2012}, and DESI surveys \citep{DESCollab2016}, matched to multi-wavelength photometry from SDSS, the DESI Legacy Imaging Surveys \citep[DESI-LS;][]{Dey2019}, WISE \citep{WISE2010}, and \textit{Gaia} \citep{deBruijne2012GaiaMission}. We further performed feature selection guided by Shapley Additive exPlanations (SHAP; \citealt{Lundberg2017,Lundberg2020}), evaluated the performance with both point-estimate and distribution-based metrics, and applied the models to both members of all systems in the MGQPC catalogue to deliver a photo-$z$ and PDF catalogue optimised for quasar pair search and follow-up.

This paper is organised as follows. We describe the photometric and spectroscopic data sets in Sect.~2. The ML approach adopted for photo-$z$ estimation is detailed in Sect.~3, and the evaluation strategy, including SHAP-based feature selection, is presented in Sect.~4. The results, including the application to the MGQPC catalogue, are discussed in Sect.~5. We conclude and outline future prospects in Sect.~6. Throughout this paper, we adopt a flat $\Lambda$ cold dark matter cosmology with $\Omega_{\Lambda} = 0.7$, $\Omega_{\rm m} = 0.3$, and $H_{0} = 70\ \mathrm{km\ s^{-1}\ Mpc^{-1}}$.

\section{Data} \label{sec2}

\subsection{Photometric data} \label{sec2.1}
The DESI-LS provides the optical and infrared imaging data to build target catalogues for DESI spectroscopic surveys. It integrates three imaging programmes: the Dark Energy Camera Legacy Survey (DECaLS), the Beijing-Arizona Sky Survey (BASS), and the Mayall $z$-band Legacy Survey (MzLS). This survey offers homogeneous photometry over a footprint of order $1.4\times10^{4}$\,deg$^{2}$ in the $g$, $r$, and $z$ bands, complemented by NEOWISE infrared photometry in the $W1$, $W2$, $W3$, and $W4$ bands \citep{Mainzer2014}. In this work we used the tenth data release of the DESI Legacy Imaging Surveys (hereafter, LS10), which expands the imaging footprint to more than $2.0\times10^4$ deg$^2$ \citep{Dey2019}. The southern LS10 footprint covers the southern extragalactic sky and includes, among others, additional DECam imaging associated with the DeROSITAS programme \citep{Saxena2024,Zenteno2025}. LS10 provides optical imaging in the $g$, $r$, $i$, and $z$ bands, with the $i$-band coverage mainly available in the southern footprint.
It also provides aperture and model fluxes, along with morphological classifications for each source. We converted the $g$, $r$, $i$, $z$, $W1$, and $W2$ bands' model and aperture fluxes into AB magnitudes. We then applied the DESI magnitude limits ($g \leqslant 24.0$, $r \leqslant 23.4$, $z \leqslant 22.5$) and removed sources with missing measurements in any required band. The remaining objects constitute our photometric sample.

The SDSS provides uniform, deep five-band ($u,g,r,i,z$) optical imaging over $\sim$ 14\,500\,deg$^{2}$, which is widely used to select galaxy and quasar targets \citep{Gunn1998,Gunn2006,Fukugita1996,Doi2010}. We adopted the public imaging reductions from SDSS-IV (e.g. DR16; \citealt{Ahumada2020}), which provide calibrated AB photometry, astrometry, and morphology throughout the Legacy footprint. For our supplementary sample, we used SDSS photometric measurements and removed entries with missing magnitudes in any of the required bands. We then imposed a uniform $i$-band quality cut by requiring a significant detection, i.e. $i > 0$ and $i_{\mathrm{err}} < 0.2171$ (corresponding to a signal-to-noise ratio (S/N) of $> 5$), and we further discarded objects with $i < 14$ to avoid possible saturation. The remaining sources formed our SDSS-based photometric quasar sample, which complements LS10 by providing $u$-band coverage and an independent measurement in the $i$ band.

\subsection{Spectroscopic data} \label{sec2.2}
We used spectroscopic redshifts as the reference targets for model training. The spectroscopic quasar sample was compiled from three resources: SDSS DR18 \citep{Almeida2023} \footnote{\url{https://dr18.sdss.org/optical/spectrum/search}}, LAMOST DR12 v1.0\footnote{\url{http://www.lamost.org/dr12/v1.0/}}, and DESI DR1\footnote{\url{https://data.desi.lbl.gov/doc/access/}}. After applying the quality criteria of each catalogue (see Table~\ref{tab:kst_surveys}), we removed quasars with problematic spectroscopic flags or non-positive redshifts ($z \leqslant 0$). The remaining sources formed our quasar sample with secure spectroscopic redshifts. Table~\ref{tab:kst_surveys} summarises the contributing spectroscopic surveys.

To enable astrometric filtering, we constructed reference samples of spectroscopically confirmed quasars from the different surveys and cross-matched them with \textit{Gaia}~DR3 \citep{GaiaCollaboration2023DR3} to obtain the proper motions, parallaxes, and the calibrated $G_{\rm BP}$, $G$, and $G_{\rm RP}$ magnitudes. To ensure reliable astrometry, we required $G \leqslant 21$ mag. This selection yielded 793\,365 sources with \textit{Gaia} proper motion and parallax measurements. In detail, we performed two positional cross-matches with a 1\arcsec\ search radius between the spectroscopic redshift sample and LS10, and between the spectroscopic redshift sample and SDSS DR16, and we obtained the nearest counterpart in each case. After applying the photometric selection cuts described in Sect.~2.1 and removing objects with missing photometry in any required band, the SDSS-matched and LS10-matched samples contained 673\,539 and 347\,824 sources, respectively. We denoted these two photometric matches of the same spectroscopic quasar sample as KSTS and KSTD, and used them for training, validation, and testing. The smaller size of KSTD was mainly driven by the substantial fraction of sources without reliable $i$-band measurements in LS10, which were excluded by our uniform $i$-band quality requirement.

Before using magnitudes and colours in the model, we de-reddened all photometry to correct the interstellar extinction. For the five SDSS bands, we corrected the photometry for galactic extinction using the dust map of \citet[][hereafter SFD98]{Schlegel1998} and the extinction coefficients for the SDSS bands to obtain the extinction-corrected AB magnitudes. For \textit{Gaia} DR3, we adopted the EDR3 passbands from \cite{Riello2021} and used the fixed extinction coefficients of $R_{G_{\rm BP}}=3.4751$, $R_G=2.8582$, and $R_{G_{\rm RP}}=1.8755$. These coefficients were computed as $R_\lambda=(A_\lambda/A_V)\,R_V$ using the optical-to-mid-IR extinction law of \citet{Wang2019} with $R_V=3.1$. 
For LS10 and the associated \textit{WISE} photometry ($g,r,i,z,W1,W2$), galactic extinction corrections were applied at the flux level using the catalogue-provided transmission factors \texttt{mw\_transmission\_band}, i.e. \texttt{flux\_band}/\texttt{mw\_transmission\_band}, before converting the corrected fluxes to AB magnitudes. Table~\ref{tab:elg_phot_extended} lists all parameters used in both KSTS and KSTD.

\begin{figure}
    \centering
    \includegraphics[width=0.8\linewidth]{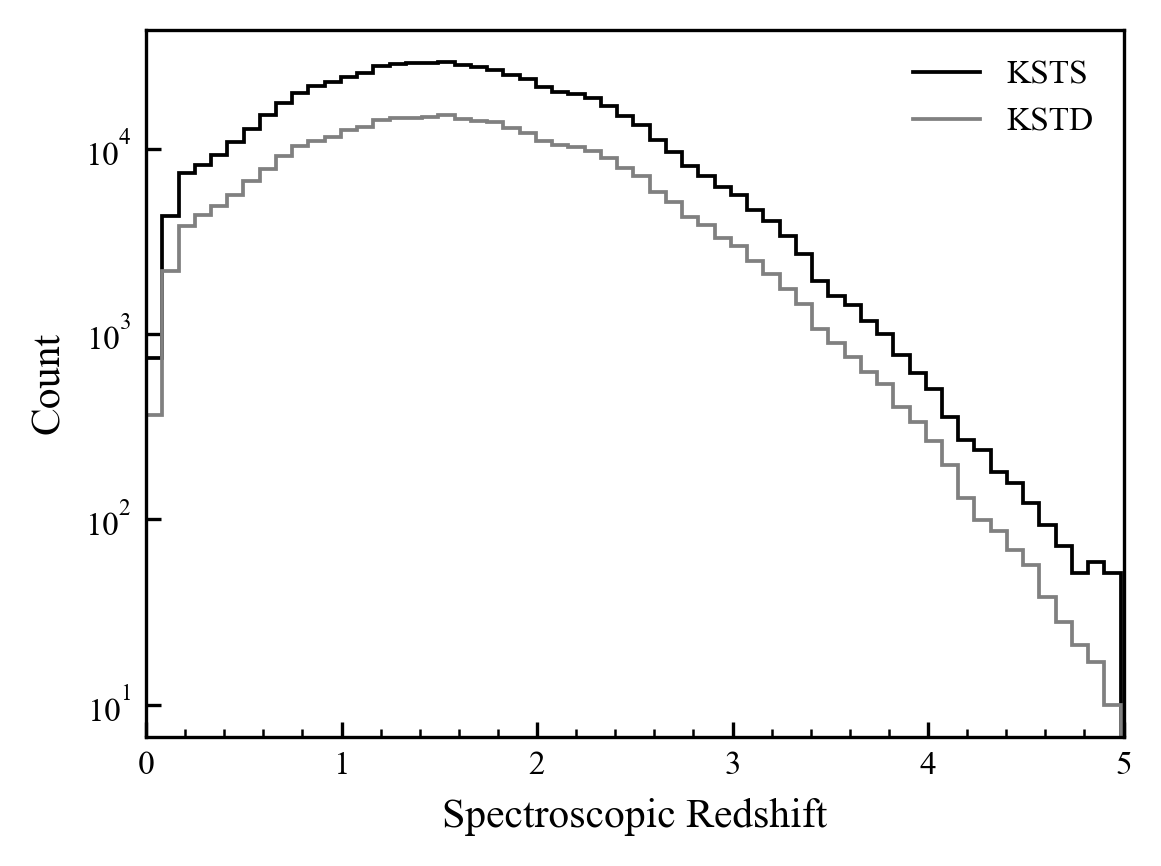}
    \caption{
    Spectroscopic redshift distributions for the KSTS (black) and KSTD (grey) quasar samples used in this work.
    The histograms are shown as step curves, and the vertical axis indicates the number of objects in each redshift bin on a logarithmic scale.
    }
    \label{fig:redshift_hist}
\end{figure}

We summarise the selection steps and sample sizes in Table~\ref{tab:spec_sample_flow}. Starting from the complete set of secure spectroscopic quasars (1\,887\,006 objects after quality cuts across SDSS~DR18, LAMOST~DR12, and DESI~DR1), we first selected the set with reliable \textit{Gaia} astrometry and then constructed the KSTS and KSTD samples requiring SDSS and LS10 imaging and good-quality photometry. The resulting KSTS and KSTD samples formed the basis for all subsequent modelling. Figure~\ref{fig:redshift_hist} presents their spectroscopic-redshift distributions.

For both KSTS and KSTD, we split the known quasar sample into three mutually exclusive sets in a fixed 3:1:1 ratio (60\,\%:20\,\%:20\,\%) for training, validation, and testing, respectively. The split was performed randomly, but stratified in redshift so that all three sets shared similar $z$ distributions. The training set was used to fit the model parameters, whereas hyperparameters were tuned and features were selected by optimising performance on the validation set. The test set was kept completely untouched during model development and was used only once, at the end of the analysis, to provide an unbiased estimate of the performance on unseen data. Unless otherwise stated, the global performance metrics quoted for KSTS and KSTD referred to either the validation or the test set, as explicitly indicated in the text.

\begin{table}
    \caption{Information on three spectroscopic surveys.}
    \centering
    \begin{tabular}{l r l c }
    \hline\hline
    Survey & No. of Quasars & Quality Criterion & $\tilde{z}_{\mathrm{spec}}$ \\
    \hline
    SDSS DR18   & 962\,830   &
      \begin{tabular}[c]{@{}l@{}}
        \texttt{class} = QSO \\
        \texttt{zWarning} = 0 \\
        \texttt{ZERR} $< 0.01$
      \end{tabular} & 1.757 \\
    LAMOST DR12 & 50\,036    &
      \begin{tabular}[c]{@{}l@{}}
        \texttt{class} = QSO \\
        \texttt{Z\_ERR} $\leqslant 0.01$
      \end{tabular} & 1.359 \\
    DESI DR1    & 1\,435\,798 &
      \begin{tabular}[c]{@{}l@{}}
        \texttt{class} = QSO \\
        \texttt{zWarning} = 0 \\
        \texttt{ZERR} $< 0.01$
      \end{tabular} & 1.752 \\
    \hline
    Total No. & 1\,887\,006 &  & 1.729\\
    \hline
    \end{tabular}
    \par\vspace*{12pt} 
    \begin{minipage}{\columnwidth}
        \footnotesize{\textbf{Notes:} $\tilde{z}_{\mathrm{spec}}$ denotes the median spectroscopic redshift.}
    \end{minipage}
    \label{tab:kst_surveys}
\end{table}

\begin{table}
    \caption{Construction of the spectroscopic samples (KSTS and KSTD).}
    \label{tab:spec_sample_flow}
    \centering
    \begin{tabular}{l r}
        \hline\hline
        Selection Step & No. of Sources \\
        \hline
        Spec-$z$ QSOs (quality cuts)\tablefootmark{a} & 1\,887\,006 \\
        With \textit{Gaia}~DR3, $G \leqslant 21$            &   793\,365 \\
        KSTS (SDSS mag cuts)                    &   673\,539 \\
        KSTD (LS10 mag cuts)                &   347\,824 \\
        KSTS: training                                   &   404\,123 \\
        KSTS: validation                                     &   134\,708 \\
        KSTS: test                                    &   134\,708 \\
        KSTD: training                                   &   208\,694 \\
        KSTD: validation                                    &    69\,565 \\
        KSTD: test                                    &    69\,565 \\
        \hline
    \end{tabular}
    \tablefoot{
        \tablefoottext{a}{Total number of quasars from SDSS~DR18, LAMOST~DR12, and DESI~DR1
        after applying the spectroscopic-quality and redshift cuts listed in Table~\ref{tab:kst_surveys}.}
    }
\end{table}

\begin{table*}
    \caption{Photometric parameters used for the regressor construction and test, including SDSS, LS10, \textit{WISE}, and \textit{Gaia} photometry.}
    \centering
    \begin{tabular}{l l c c}
    \hline \hline
     Parameter & Definition & Catalogue & Waveband \\
    \hline
    $u$  & Model magnitude in $u$-band & SDSS & Optical band \\
    $g$  & Model magnitude in $g$-band & SDSS & Optical band \\
    $r$  & Model magnitude in $r$-band & SDSS & Optical band \\
    $i$  & Model magnitude in $i$-band & SDSS & Optical band \\
    $z$  & Model magnitude in $z$-band & SDSS & Optical band \\
    \hline
    $G_{\rm BP}$ & Model magnitude in $BP$-band & \emph{Gaia} & Optical band \\
    $G$  & Model magnitude in $G$-band  & \emph{Gaia} & Optical band \\
    $G_{\rm RP}$ & Model magnitude in $RP$-band & \emph{Gaia} & Optical band \\
    \hline
    $g$  & Model magnitude in $g$-band & LS10 & Optical band \\
    $r$  & Model magnitude in $r$-band & LS10 & Optical band \\
    $i$  & Model magnitude in $i$-band & LS10 & Optical band \\
    $z$  & Model magnitude in $z$-band & LS10 & Optical band \\
    $g\_ap$(1--8)\textsuperscript{a} & Aperture magnitude in $g$-band & LS10 & Optical band \\
    $r\_ap$(1--8) & Aperture magnitude in $r$-band & LS10 & Optical band \\
    $i\_ap$(1--8) & Aperture magnitude in $i$-band & LS10 & Optical band \\
    $z\_ap$(1--8) & Aperture magnitude in $z$-band & LS10 & Optical band \\
    type & Morphological type flag  & LS10 & Optical band \\
    \hline
    $W1$ & Model magnitude in $W1$-band & \textit{WISE} & Infrared band \\
    $W2$ & Model magnitude in $W2$-band & \textit{WISE} & Infrared band \\
    $W1\_ap$(1--5)\textsuperscript{b} & Aperture magnitude in $W1$-band &\textit{WISE} & Infrared band \\
    $W2\_ap$(1--5) & Aperture magnitude in $W2$-band & \textit{WISE} & Infrared band \\
    \hline  
    z\_spec & Spectroscopic redshift & Known sample & - \\
    \hline
    \end{tabular}
    \begin{minipage}{\linewidth}
    \footnotesize
    \vspace{3pt}
    \textbf{Notes.}
    \textsuperscript{a}\,Optical aperture magnitude suffixes 1--8 correspond to fluxes measured within circular apertures of radii of 0.5, 0.75, 1.0, 1.5, 2.0, 3.5, 5.0, and 7.0 arcsec.\\
    \textsuperscript{b}\,Infrared aperture magnitude suffixes 1--5 correspond to apertures with radii of 3, 5, 7, 9, and 11 arcsec.
    \end{minipage}
    \label{tab:elg_phot_extended}
\end{table*}

\section{Method} \label{sec3}

\subsection{CatBoost} \label{sec3.1}
\textsc{CatBoost} is an efficient implementation of gradient-boosted decision trees (GBDTs; \citealt{Friedman2001}) designed to handle categorical (and text-derived) features while avoiding prediction shift. It combines two key ideas: ordered boosting, whereby each example is trained using only prefix sets under random permutations, and ordered target statistics, which encode categorical variables via leak-free, prior-smoothed target estimates computed on disjoint prefixes \citep{Dorogush2018,Prokhorenkova2018}. As base learners, \textsc{CatBoost} employs symmetric (oblivious) trees in which all nodes at the same depth share the same split; this structure reduces variance, simplifies regularisation, and yields cache-friendly, highly parallel inference on CPUs and GPUs. In practice, \textsc{CatBoost} offers strong out-of-the-box performance with limited hyperparameter tuning, native support for missing values, and scalable training for regression, classification, and ranking tasks. In this work, we leverage these properties to achieve a stable generalisation of heterogeneous tabular data with mixed numerical and categorical input.

\subsection{FlexZBoost} \label{sec3.2}
\textsc{FlexZBoost} \citep{IzbickiLee2017,Dalmasso2020} is a CDE method of photo-$z$ estimation built on the \textsc{FlexZBoost} framework\footnote{\url{https://github.com/lee-group-cmu/FlexCode}}. It represents the conditional probability density, $p(z \mid \boldsymbol{x})$, as an expansion in an orthonormal basis and estimates the associated coefficients by regressing the evaluations of the basis functions at the target redshift on the covariates. A commonly used instantiation employs the gradient-boosted trees (\textsc{XGBoost}; \citealt{ChenGuestrin2016}) to fit these regressions, while the model complexity (e.g. the number of basis terms and related hyperparameters) is selected by minimising a validation CDE loss \citep{Dalmasso2020}. The framework is regressor-agnostic. In principle, any ML model that can regress the basis-expansion coefficients can be used, and the method returns full photo-$z$ PDFs from which point summaries (e.g. the mode or mean) and the calibrated uncertainty measures can be derived. On LSST-like mock benchmarks, \textsc{FlexZBoost} achieved the lowest CDE loss among the methods evaluated by \citet{Schmidt2020}, within that experimental set-up.

\section{Experiment} \label{sec4}
We trained the models on the training set and tuned the model settings using the validation set, which was also used for early stopping. The test set was kept completely independent, and used only for the final performance evaluation on previously unseen data.

\subsection{Metrics} \label{sec4.1}
We evaluated the model performance on photo-$z$ estimation using two complementary categories of metrics. The first one was the distribution-to-point metrics, which assess the full predictive distribution by evaluating it at the true spectroscopic redshift and thereby probe the calibration and sharpness. The second set was a suite of point-to-point metrics that compare scalar summaries of the predictions with the spectroscopic values, quantifying the accuracy and robustness \citep{RAIL2025}. This comprehensive evaluation examines both the statistical reliability and the accuracy of our predictions. Following the common practice in the photo-$z$ literature, we adopted a standard set of regression metrics to evaluate the accuracy and reliability of photo-$z$. $z_{\rm spec}$ and $z_{\rm phot}$ are used to denote the spectroscopic (true) and photometric (predicted) redshifts, respectively. Formal definitions are provided below:

\begin{itemize}
\item Normalised redshift residual (e.g. \citealt{Cohen2000,Ilbert2006}). 
It's a scale-free residual. Those values near zero indicate low bias after accounting for redshift stretching.
\begin{equation}
\Delta z \;=\; \frac{z_{\rm spec}-z_{\rm phot}}{1+z_{\rm spec}} .
\end{equation}

\item RMSE (root mean squared error). 
It is used to capture the overall dispersion of the absolute errors, and is particularly sensitive to the outliers (lower is better).
\begin{equation}
\sigma_{\rm RMSE} \;=\; \sqrt{\frac{1}{N}\sum_{i=1}^{N}\left(z_{{\rm spec},i}-z_{{\rm phot},i}\right)^{2}} . 
\end{equation}

\item Normalised median absolute deviation \citep[NMAD; ][]{Brammer2008,Ilbert2006,Salvato2019}. 
It is a robust scatter measure and less sensitive to the outliers than the RMSE (lower is better).
\begin{equation}
\sigma_{\rm NMAD} \;=\; 1.4826 \times \operatorname{median}\!\left(\left\lvert\Delta z \right\rvert\right) . 
\end{equation}

\item Outlier fraction (conventional cut-off $\lvert\Delta z\rvert>0.15$; \citealt{Hildebrandt2012}). 
It is the fraction of the catastrophic deviations beyond the threshold (lower is better).
\begin{equation}
\eta \;=\; \frac{N\!\left(\left\lvert\Delta z\right\rvert>0.15\right)}{N_{\rm total}} . 
\end{equation}

\item Probability integral transform \citep[PIT; ][]{Polsterer2016,zhao2021,Dey2025LADaR}. 
It is a calibration diagnostic. For the well-calibrated PDFs, the distribution of $u$ can be approximated by a uniform distribution over the interval $[0,1]$.
\begin{equation}
u \;=\; \mathrm{CDF}\!\left(z_{\rm spec}\right) \;=\; \int_{0}^{z_{\rm spec}} \mathrm{PDF}(z)\,dz \;\in [0,1] . 
\end{equation}
\end{itemize}

\subsection{Feature selection} \label{sec4.2}

\begin{figure}
    \centering
    \includegraphics[width=0.48\textwidth]{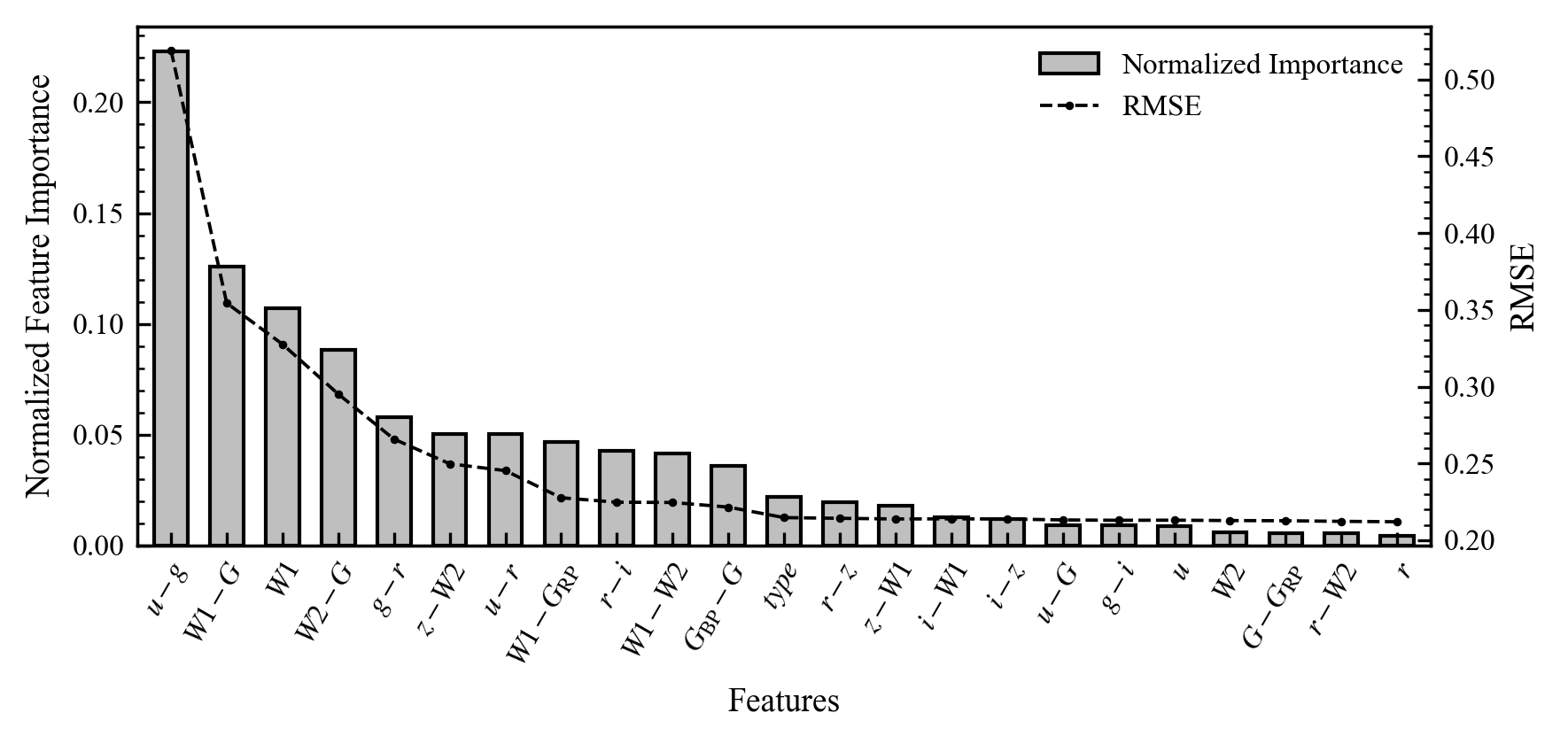}
    \caption{Feature importance and validation RMSE for the KSTS sample. Normalised SHAP importance (left axis) of each feature is shown by bars. The dashed line represents the validation RMSE (right axis) achieved by cumulatively adding features in descending order of SHAP importance. For clarity, only the top 23 most important features are displayed.    
    }
    \label{fig:fs_shap_rfe}
\end{figure}

We assessed and selected the input variables using a SHAP-guided, stage-wise procedure based on the recursive feature elimination (RFE). We began with a set of photometric features that includes both model-based and aperture-based measurements. A baseline \textsc{CatBoost} regressor was trained with five-fold cross-validation, and the mean absolute SHAP value of each feature was computed on the validation part of each fold \citep{Lundberg2020}. We aggregated the SHAP values across folds to obtain a global ranking; for presentation purposes only, the scores were normalised to sum to unity. We then ran RFE based on this ranking. Starting from the full set, we iteratively removed the feature with the smallest median absolute SHAP value, retrained the model, and tracked the validation RMSE as a function of the retained feature number, $k$ (Fig.~\ref{fig:fs_shap_rfe}). We adopted $k_{\rm opt}$ as the value that minimises the validation RMSE, and monitored $\sigma_{\rm NMAD}$ and the outlier fraction as the secondary diagnostics. This procedure yielded a minimum validation RMSE at $k_{\rm opt}=23$ for KSTS, and $k_{\rm opt}=53$ for KSTD. In both cases, the selected features were dominated by broad-band colours and fluxes. These observables capture both the redshift-dependent continuum shape and the imprint of prominent emission lines moving through the filters, while the mid-infrared colours trace the characteristic hot-dust excess of quasars. Together, they help mitigate redshift degeneracies in broad-band photometry, driven by both continuum-colour similarities and the aliasing introduced as prominent emission lines move through the filters. The final feature lists for the KSTS and KSTD samples are provided in Appendix~\ref{app:features}.

For completeness, we tested \textsc{CatBoost}’s built-in importance. The top features largely agreed with the SHAP order, but the rankings of the correlated, low-salience variables were less stable across folds. We therefore treated these measures as reference only: they do not drive selection, and we omitted their plots and tables. Instead, we relied on SHAP, which provides per-feature contributions with their signs, making the feature effects easier to be interpreted and more stable when inputs are correlated. This workflow is inspired by the feature-selection strategy of \citet{Li2022} and specifically designed to improve both stability and interpretability.

\begin{figure*}
    \centering
    \includegraphics[width=0.72\textwidth]{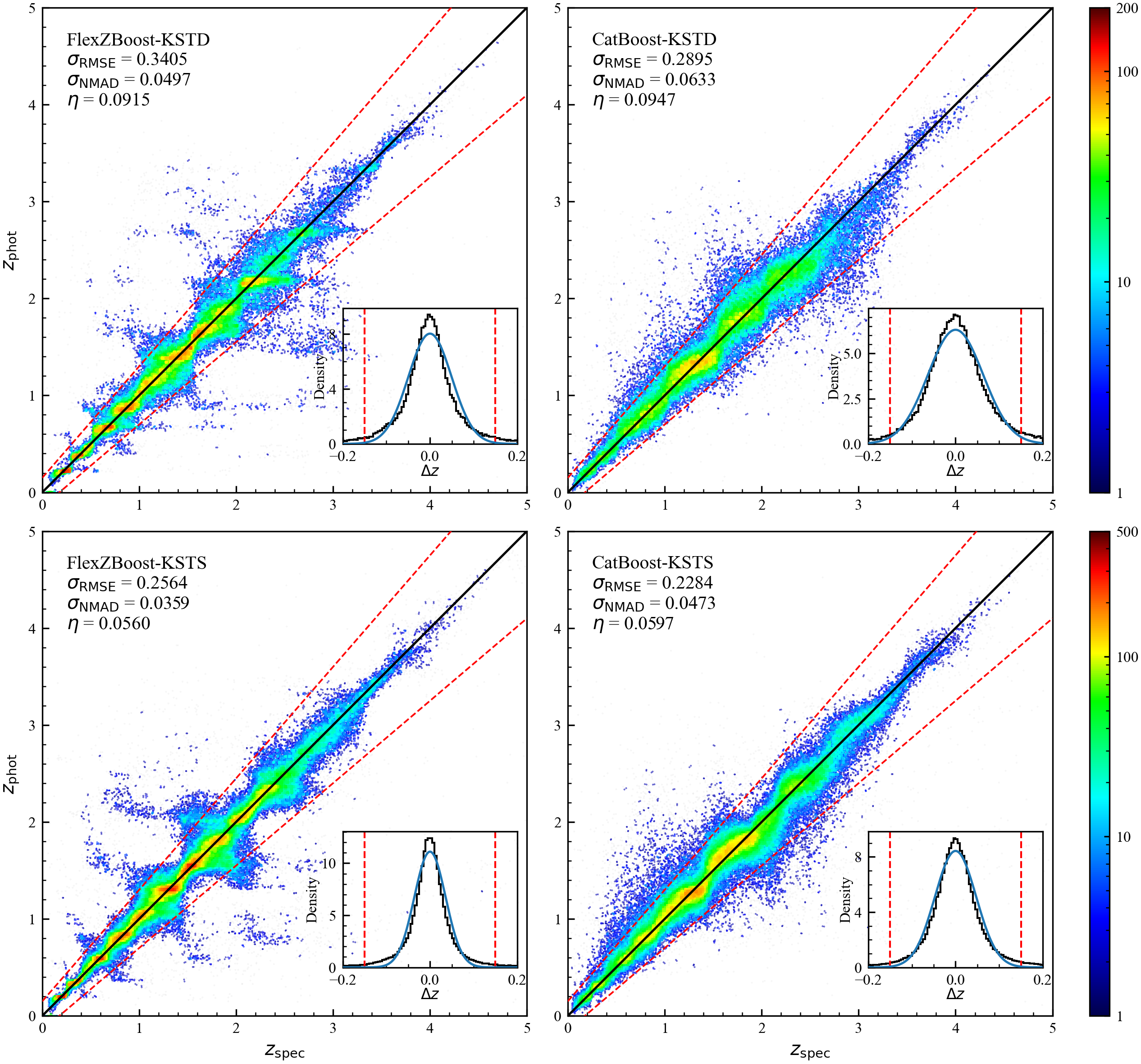}
    \caption{Photo-$z$ performance for two algorithms (\textsc{FlexZBoost}, left; \textsc{CatBoost}, right) on two samples (KSTD, top; KSTS, bottom). In each big panel, the scatter plot shows $z_{\rm phot}$ versus $z_{\rm spec}$. The solid line is the identity relation; dashed lines ($\lvert\Delta z\rvert=0.15$) mark the outlier threshold, where $\Delta z=(z_{\rm spec}-z_{\rm phot})/(1+z_{\rm spec})$. Insets list $\sigma_{\rm RMSE}$, $\sigma_{\rm NMAD}$, and the outlier fraction, $\eta$. The small panels display the distribution of $\Delta z$ as black histograms; the vertical dashed red lines mark the outlier threshold at $\pm 0.15$, and the solid blue curves show Gaussian profiles with widths corresponding to $\sigma_{\rm NMAD}$.}
    \label{fig:zp-zs}
\end{figure*}

\section{Results and discussion} \label{sec5}

\subsection{Results} \label{sec5.1}
We quantified the performance of our photo-$z$ models on the test sets (see Table~\ref{tab:photoz_subsamples_ksts_kstd}). Overall, the results on KSTS are slightly better than those on KSTD. This difference is unlikely to arise from a single factor alone. Among the plausible contributors, the $u$ band is expected to play an important role because it anchors the blue end of the SED and adds strong colour leverage through quantities such as $u-g$. For the low- to intermediate-redshift quasars, it also captures the UV-excess behaviour that changes with redshift. At $z \lesssim 1$, the $u$-band photometry further helps constrain low-$z$ spectral breaks, including the 4000\,\AA\ break when host galaxy light becomes non-negligible. Without the $u$ band, the remaining bands leave stronger degeneracies among the continuum slope, spectral features, and emission line contributions. This is expected to increase both the scatter and the catastrophic outlier rate, although survey-dependent photometric systematics may also contribute to the observed KSTS--KSTD difference. The test sets are independent of training and validation sets, so the reported metrics estimate the performance on previously unseen data. In Sect.~5.4 we also found comparable performance on the external MGQPC\_A sub-sample with reliable spectra, which provides an independent check of the generalisation to the MGQPC application domain.

To disentangle the role of the SDSS $u$ band from possible bright-source systematics in LS10, we carried out two control tests (Table~\ref{tab:catboost_control_tests}). First, we repeated the KSTS experiment after removing the $u$-band information. The performance degrades substantially, with $\sigma_{\rm NMAD}$ increasing from 0.047 to 0.060, $\sigma_{\rm RMSE}$ from 0.228 to 0.308, and the outlier fraction, $\eta$, from 5.968\% to 10.683\%. Second, we repeated the KSTD experiment after excluding sources affected by saturation-related flags, and then after an additional contamination-filtering step (i.e. excluding sources whose LS10 optical photometry shows stronger masking or flux contamination, or a lower fraction of in-source flux in the $griz$ bands). In contrast, the resulting changes are comparatively modest: the KSTD test performance changes from $\sigma_{\rm NMAD}=0.063$, $\sigma_{\rm RMSE}=0.290$, and $\eta=9.473\%$ in the original sample to 0.062, 0.285, and 9.234\% after saturation cleaning, and to 0.062, 0.284, and 9.061\% after the additional contamination filtering. Because the cleaning steps change the KSTD sample composition and size, this comparison should be interpreted as indicative of the relative importance of these effects rather than as a strictly one-to-one differential test. These control tests therefore suggest that the inclusion of the SDSS $u$ band is likely the main driver of the better KSTS performance, while saturation-related bright-source photometric systematics in LS10 are plausibly a secondary contributor rather than the sole explanation.

Fig.~\ref{fig:zp-zs} compares the photometric and spectroscopic redshifts for the two algorithms and the two samples: \textsc{FlexZBoost} (left) and \textsc{CatBoost} (right) on KSTD (top) and KSTS (bottom). In all big panels, the binned density scatter clusters tightly around the $z_{\rm phot}=z_{\rm spec}$ line, with a visibly lower outlier fraction in KSTS than in KSTD, particularly for the PSF sources. Consistent with our SHAP-guided salience analysis (Fig.~\ref{fig:fs_shap_rfe}), the features encoding blue-end leverage (including $u$-band information) and size-sensitive aperture contrasts rank among the top contributors; accordingly, KSTS exhibits tighter clustering and fewer outliers than KSTD. The solid blue curves overplotted in the inset panels show Gaussian profiles with widths set by the corresponding $\sigma_{\rm NMAD}$ values, allowing for a direct comparison of the residual distributions among the four cases.

\begin{figure}
    \centering
   \includegraphics[width=0.48\textwidth]{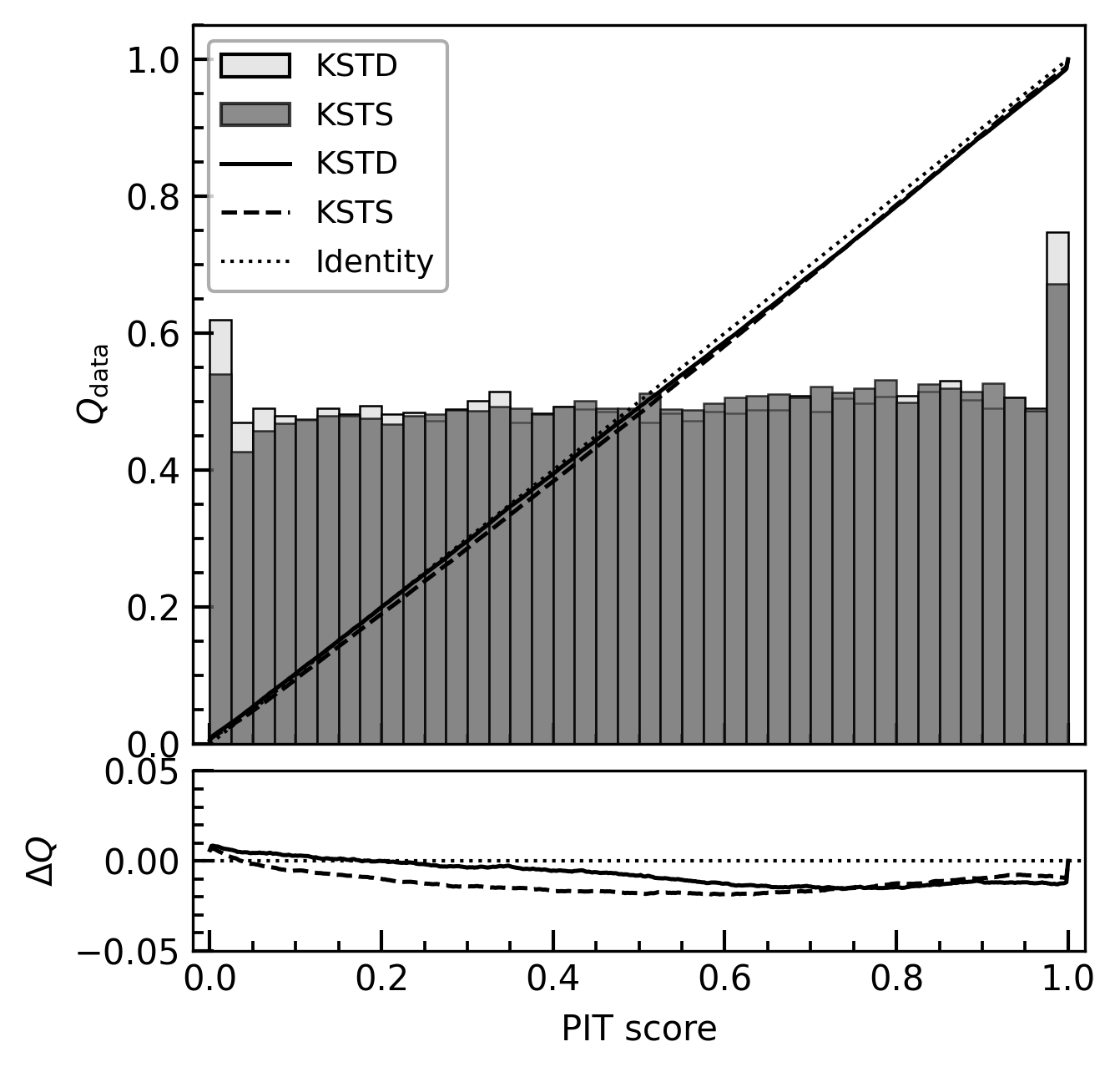}
    \caption{Calibration diagnostics for the redshift PDFs. The top panel displays the PIT-QQ plot for the test samples, comparing the identity line (for a flat histogram under perfect calibration) with the PIT values from our redshift PDFs for KSTD (open bars, solid curve) and KSTS (filled bars, dashed curve). The bottom panel shows the difference from identity ($\Delta Q$), highlighting any systematic biases or trends between the two samples.}
    \label{fig:pitqq}
\end{figure}

We assessed the global calibration of the predicted redshift PDFs using the PIT \citep{Dawid1984,Gneiting2005}, a widely adopted diagnostic in this context \citep{Pasquet2018,Schuldt2021,NewmanGruen2022}. Under ideal calibration, the PIT values are uniformly distributed (i.e. a flat histogram); any departure from uniformity indicates miscalibration. Specifically, a U-shaped PIT histogram indicates underdispersion (undercoverage): the predicted PDFs are too narrow and their credible intervals do not attain the nominal coverage. On the other hand, an inverted-U (hump-shaped) PIT indicates overdispersion (overcoverage): the PDFs are too broad and the credible intervals attain coverage above the nominal level \citep{DIsanto2018}. The spikes near 0 or 1 typically indicate catastrophic outliers, where the true redshift falls in the extreme tails of the predicted distribution and is assigned a very low probability density.

Fig.~\ref{fig:pitqq} presents the calibration of the predicted photo-$z$ PDFs using PIT histograms (top panel) and PIT quantile-quantile (QQ) plots (bottom panel). We compared our diagnostics qualitatively to Fig.~9 of \citet{Roster2024PICZL}, which presents similar tests for active galactic nuclei (AGNs). The QQ plots show the empirical cumulative distribution function of the PIT values, denoted $Q_{\rm data}(x)$, against the uniform expectation $U(0,1)$. Under ideal calibration, $Q_{\rm data}(x)$ lies on the 1:1 line. Both the results based on KSTS and KSTD test samples remain close to this line, indicating near-uniform PIT values and good global calibration. The KSTS QQ curve shows slightly smaller deviations from the 1:1 line. The remaining differences are modest and suggest that the predicted PDFs are slightly too narrow, with mildly underestimated uncertainties. We emphasise that the PIT is used here purely as a diagnostic.

\begin{table*}
    \caption{Comparison of bias, scatter, RMSE, and outlier fraction for the KSTS and KSTD test samples and the MGQPC\_A sub-sample.}
    \centering
    \begin{tabular}{l l l r r r r r}
    \hline\hline
    Method & Survey & Sample & No. of Sources & $\Delta z$ & $\sigma_{\rm NMAD}$ & $\sigma_{\rm RMSE}$ & $\eta$ [\%] \\
    \hline
    \textsc{FlexZBoost} & KSTS & Test        & 134\,708 & $-0.004$ & 0.036 & 0.256 & 5.600 \\
                       &      & Type: PSF   & 125\,776 & $-0.004$ & 0.036 & 0.261 & 5.786 \\
                       &      & Type: EXT   & 8\,932   & $0.001$  & 0.032 & 0.182 & 2.978 \\
                       &      & S/N $>$ 3   & 122\,470 & $-0.003$ & 0.035 & 0.238 & 5.200 \\
                       &      & S/N $<$ 3   & 12\,238  & $-0.010$ & 0.044 & 0.395 & 9.601 \\
                       &      & MGQPC\_A    & 2\,576   & $-0.002$ & 0.028 & 0.294 & 5.124 \\[3pt]
                       & KSTD & Test        & 69\,565  & $-0.001$ & 0.050 & 0.341 & 9.147 \\
                       &      & Type: PSF   & 64\,775  & $-0.002$ & 0.051 & 0.347 & 9.538 \\
                       &      & Type: EXT   & 4\,790   & $0.007$  & 0.038 & 0.236 & 3.862 \\
                       &      & S/N $>$ 3   & 66\,687  & $-0.002$ & 0.049 & 0.337 & 8.967 \\
                       &      & S/N $<$ 3   & 2\,878   & $0.001$  & 0.057 & 0.418 & 13.308 \\
                       &      & MGQPC\_A    & 1\,063   & $0.002$  & 0.029 & 0.351 & 7.714 \\
    \hline
    \textsc{CatBoost}   & KSTS & Test        & 134\,708 & $-0.007$ & 0.047 & 0.228 & 5.968 \\
                       &      & Type: PSF   & 125\,879 & $-0.007$ & 0.048 & 0.231 & 6.074 \\
                       &      & Type: EXT   & 8\,829   & $-0.006$ & 0.039 & 0.191 & 4.451 \\
                       &      & S/N $>$ 3   & 122\,342 & $-0.006$ & 0.047 & 0.215 & 5.520 \\
                       &      & S/N $<$ 3   & 12\,366  & $-0.016$ & 0.052 & 0.331 & 10.399 \\
                       &      & MGQPC\_A    & 2\,576   & $-0.007$ & 0.043 & 0.270 & 6.289 \\[3pt]
                       & KSTD & Test        & 69\,565  & $-0.013$ & 0.063 & 0.290 & 9.473 \\
                       &      & Type: PSF   & 64\,735  & $-0.013$ & 0.065 & 0.295 & 9.832 \\
                       &      & Type: EXT   & 4\,830   & $-0.006$ & 0.045 & 0.196 & 4.658 \\
                       &      & S/N $>$ 3   & 66\,644  & $-0.012$ & 0.063 & 0.286 & 9.305 \\
                       &      & S/N $<$ 3   & 2\,921   & $-0.020$ & 0.073 & 0.354 & 13.317 \\
                       &      & MGQPC\_A    & 1\,063   & $-0.001$ & 0.053 & 0.293 & 8.467 \\
    \hline
    \end{tabular}
    \par\vspace*{12pt}
    \begin{minipage}{\linewidth}
    \footnotesize
    \textbf{Notes.}
    Sub-samples: (1) Test sample includes all sources in the corresponding KSTS or KSTD test set;
    (2) Type: PSF/EXT split point-like and extended sources according to the \texttt{type} morphological classifier;
    (3) S/N $>$ 3 and S/N $<$ 3 sub-samples are defined using the minimum S/N among all available bands;
    (4) MGQPC\_A denotes the set of MGQPC components with reliable spectroscopic redshifts and clean photometry, used as an external test sample.
    \end{minipage}
    \label{tab:photoz_subsamples_ksts_kstd}
\end{table*}

\begin{figure}
    \centering
    \includegraphics[width=1.0\linewidth]{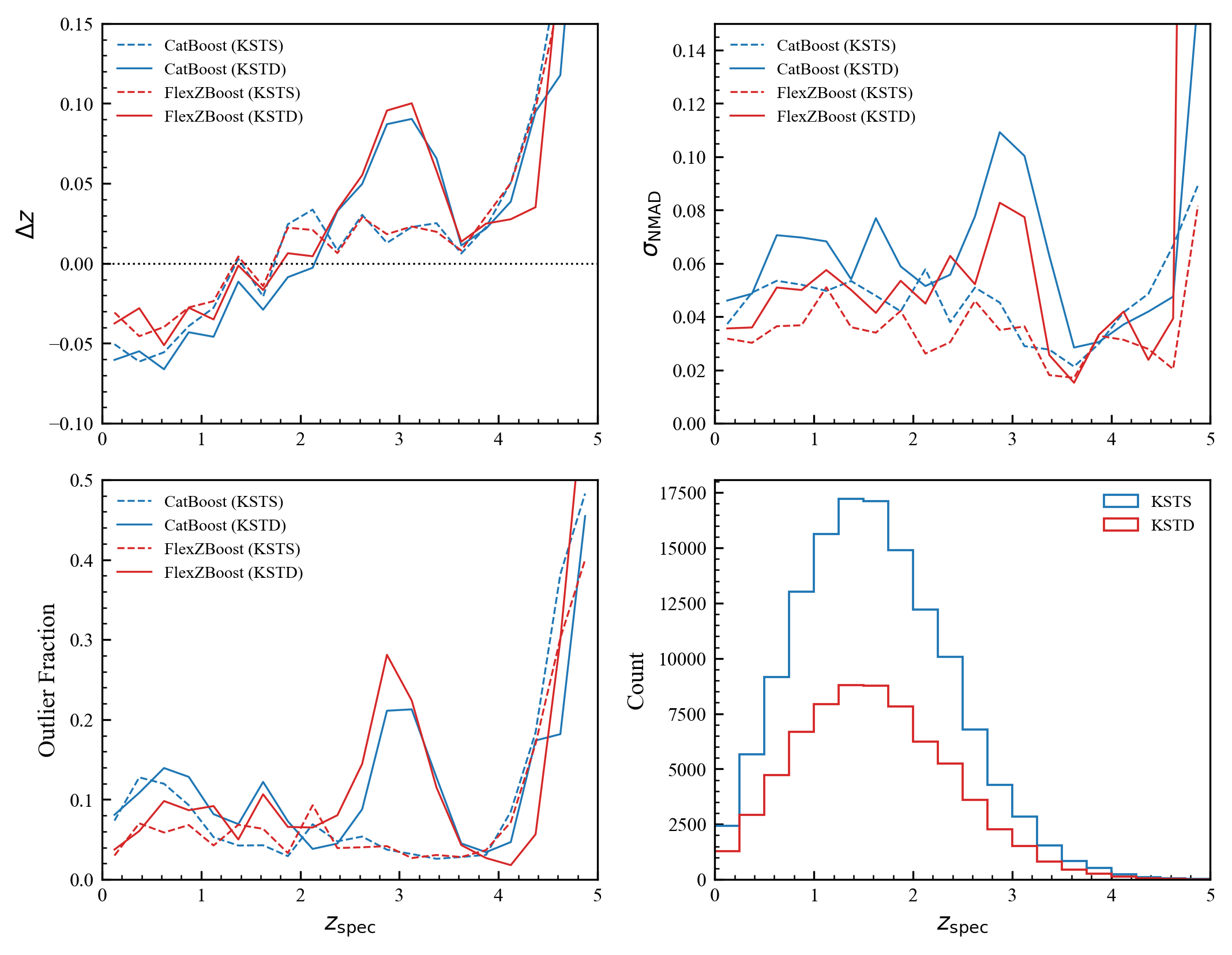}
    \caption{
    Redshift-binned photo-$z$ performance for \textsc{CatBoost} (blue) and \textsc{FlexZBoost} (red) on the KSTD (solid lines) and KSTS (dashed lines) samples.
    From top to bottom: $\Delta z$, scatter quantified by $\sigma_{\rm NMAD}$; outlier fraction, $\eta$; and the spectroscopic redshift distribution of the two samples.
    All curves were computed in the same $z_{\rm spec}$ bins to highlight systematic trends with redshift.
    }
    \label{fig:redshift_metrics}
\end{figure}

\begin{table*}
    \caption{Control tests on the influence of the SDSS $u$-band and LS10 bright-source cleaning on \textsc{CatBoost} performance in the test sample.}
    \centering
    \begin{tabular}{l l r r r r r}
    \hline\hline
    Survey &
    Experiment &
    No. of Sources &
    $\Delta z$ &
    $\sigma_{\rm NMAD}$ &
    $\sigma_{\rm RMSE}$ &
    $\eta$ [\%] \\
    \hline
    KSTS &
    Test &
    134\,708 &
    $-0.007$ &
    0.047 &
    0.228 &
    5.968 \\
    &
    Test without $u$ band &
    134\,708 &
    $-0.012$ &
    0.060 &
    0.308 &
    10.683 \\
    \hline
    KSTD &
    Test &
    69\,565 &
    $-0.013$ &
    0.063 &
    0.290 &
    9.473 \\
    &
    Test after saturation cleaning &
    68\,339 &
    $-0.012$ &
    0.062 &
    0.285 &
    9.234 \\
    &
    Test after additional contamination filtering &
    44\,366 &
    $-0.012$ &
    0.062 &
    0.284 &
    9.061 \\
    \hline
    \end{tabular}
    \par\vspace*{12pt}
    \begin{minipage}{\linewidth}
    \footnotesize
    \textbf{Notes.}
    This table lists the full test-sample results for the control experiments discussed in Sect.~\ref{sec5.1}. The KSTS control test removes the SDSS $u$-band information, while the KSTD control tests successively exclude sources affected by saturation-related flags and by additional contamination criteria. Because the cleaning steps remove objects from the parent sample, the KSTD test-sample sizes differ across experiments.
    \end{minipage}
    \label{tab:catboost_control_tests}
\end{table*}

\subsection{Redshift trends and comparison with previous work} \label{sec5.2}
It is now well established that the quality of AGN photo-$z$ estimates is primarily controlled by the photometric quality, wavelength coverage, and representativeness of the spectroscopic training set, rather than by the choice of a specific regression algorithm \citep[e.g. ][]{Salvato2019,Newman2022,Roster2024PICZL}.
Our results support this picture. Fig.~\ref{fig:redshift_metrics} shows the redshift-binned photo-$z$ performance of \textsc{CatBoost} and \textsc{FlexZBoost} on the KSTS and KSTD samples as a function of $z_{\rm spec}$. In the redshift interval containing the bulk of our objects ($z_{\rm spec}\lesssim2$), all four combinations remain close to unbiased, with $\lvert\langle\Delta z\rangle\rvert\lesssim0.02$, $\sigma_{\rm NMAD}$ $\sim$ 0.03--0.05, and outlier fractions typically below about 10 percent. In this regime, the curves largely overlap, and the remaining differences mainly appear as modest trade-offs between $\sigma_{\rm NMAD}$ and outlier fraction. Overall, these results are broadly consistent with previous AGN photo-$z$ studies \citep{Li2022,Roster2024PICZL}, while also reflecting the specific properties of our quasar training set and photometric data. At the same time, a direct comparison with some previous studies requires caution: Li et al. \citep{Li2022} focus on unobscured AGNs, whereas Roster et al. \citep{Roster2024PICZL} consider a broader AGN sample that includes non-quasar sources. We also note several recent quasar-focused photo-$z$ studies \citep{Yao2023,Nakazono2024,Zhang2024,2025Fu}. These works provide useful quasar-specific context for the present study, although differences in photometric band sets, input modalities, redshift coverage, sample construction, and evaluation metrics limit a direct one-to-one comparison.

The systematic differences between the SDSS-based KSTS and LS10-based KSTD samples are most naturally interpreted in terms of the data rather than the models. SDSS provides uniform $u,g,r,i,z$ imaging \citep{Abazajian2009,Ahumada2020}. LS10 provides deeper and highly homogeneous $g,r,z$ photometry over a much larger area, together with all-sky $W1,W2$ infrared data \citep{Dey2019}. However, the LS10 $i$-band imaging is compiled from several DECam programmes and is not spatially complete. Our requirement of robust $g,r,i,z,W1,W2$ measurements therefore removed many LS10 sources with missing or low-quality $i$-band data, reducing both the effective size and colour coverage of the LS10 training and validation sets. In addition, the SDSS $u$ band provides strong leverage for separating $z\sim3$ quasars from lower-redshift red or dusty contaminants, as was emphasised in early AGN photo-$z$ work \citep[e.g. ][]{Ilbert2006,Salvato2011}. This is also consistent with the stronger $z_{\rm spec}\sim3$ degradation seen for KSTD compared to KSTS in Fig.~\ref{fig:redshift_metrics}, where the absence of the $u$-band constraint is expected to aggravate colour-redshift degeneracies. Our control tests (Table~\ref{tab:catboost_control_tests}) indicate that the inclusion of the SDSS $u$ band is likely the main factor, while the non-uniform $i$-band availability in LS10, saturation-related bright-source systematics, and differences in observing strategy or temporal co-addition may also contribute at a secondary level.

The trends with morphology and S/N are also consistent with previous studies. Extended sources generally show smaller scatter and fewer catastrophic outliers than point-like sources. This likely reflects the contribution of the resolved host galaxy light, which introduces additional redshift-sensitive information beyond AGN-dominated colours. This advantage may also be aided by the typically higher S/N of extended sources \citep[e.g. ][]{Cardamone2010,Salvato2011}.
Likewise, photo-$z$ estimates are significantly more accurate for objects with $\mathrm{S/N}>3$ in all bands than for those with $\mathrm{S/N}<3$, indicating that the photometric depth and measurement uncertainties are dominant drivers of the photo-$z$ performance \citep{NewmanGruen2022}. This trend also suggests that incorporating S/N-related or photometric-uncertainty features is likely to be a useful extension of the present model, especially in helping the regressor distinguish measurements of differing reliability and implicitly down-weight less reliable photometric inputs. Consistent with this, recent AGN photo-$z$ work based on LS10 has shown, through explicit feature-importance analyses, that several S/N- and error-related quantities can carry non-negligible predictive information in different source sub-samples \citep{Saxena2024}.

At higher redshifts, the performance shows clear structure. At $z$ $\sim$ 2.0--2.5, the curves begin to deviate from the low-redshift plateau. From $z$ $\sim$ 2.6--3.0, the mean residual becomes systematically positive, and both $\sigma_{\rm NMAD}$ and the outlier fraction rise to local maxima. This peak is expected in the $z\sim3$ transition regime where colour information becomes less diagnostic. With increasing redshift, the Ly$\alpha$ break at $1216\,\AA$ shifts through the $g$ band, and the Ly$\alpha$ forest suppresses the continuum bluewards of Ly$\alpha$. As the Lyman limit at $912\,\AA$ enters the $u$ band, the flux in that band becomes very faint or even non-detectable.
Prominent UV emission line complexes, such as Ly$\alpha$ and \ion{C}{iv}, also migrate across optical passband boundaries, namely the blue and red cut-offs, which increases the non-linearity of the colour redshift relation and can make it non-monotonic and degenerate. When the bluest band constraint is absent or effectively weak, most notably for the LS10-based KSTD sample that lacks the $u$-band photometry, the $z\sim3$ quasars overlap more strongly with lower redshift red or dusty contaminants, and the models will turn to favour a lower-redshift solution, consistent with the systematically positive mean residual. Additional contributions may come from larger photometric uncertainties and variability induced colour scatter.

At even higher redshifts, specifically for $z\gtrsim3$, the curves show a mild turnover near $z$ $\sim$ 3.5--4. Quasars at high redshifts gradually exhibit distinctive drop-out colours (i.e. near-non-detections in the bluest bands, yielding very red colours). However, as is shown in Fig.~\ref{fig:redshift_metrics}, the spectroscopic samples become sparse, leading to a situation in which a small number of objects can  significantly influence individual bins.
This non-monotonic behaviour at high redshifts has also been observed in other analyses of AGN photo-$z$ \citep[e.g. ][]{Salvato2011,Roster2024PICZL}.

\subsection{Limitations and systematic uncertainties} \label{sec5.3}
At the same time, several important limitations and systematic effects should be borne in mind. First, the quality and representativeness of the spectroscopic redshifts impose a strict lower limit on the achievable performance. Large AGN spectroscopic samples inevitably contain a non-negligible fraction of uncertain or incorrect redshifts, arising from ambiguous line identifications, low S/N, and occasional pipeline failures; this has been well documented for SDSS and other surveys \citep[e.g. ][]{Ahumada2020,Lyke2020}. Such errors are inherently incorporated into the measured photometric scatter and outlier fractions, especially in the high-$z$ and low-S/N bins, where even a
small number of erroneous spec-$z$ can significantly increase the apparent photo-$z$ error budget. Unfortunately, no purely photometric model can rectify these issues post hoc.

Second, quasars are intrinsically variable, and the multi-band photometry used here is generally non-simultaneous across filters. In some surveys, the reported flux in a single band may also be accumulated or stacked over long temporal baselines rather than measured as a true single-epoch flux. Variations of $\sim$ 0.1--0.3 mag, or even larger amplitudes, are commonly observed on timescales from months to years \citep[e.g. ][]{Schmidt2010,MacLeod2010,Hernitschek2016,Simm2016}. Consequently, the adopted multi-band photometry in the $u$, $g$, $r$, $i$, $z$, $W1$, and $W2$ bands may not correspond to any true instantaneous SED. This can distort the effective continuum shape and increase both the scatter and the risk of catastrophic outliers in the empirical colour-redshift relation. \citet{Simm2015} showed explicitly that, for variable AGNs, photo-$z$ performance is best when the adopted multi-band photometry approximates a snapshot SED, whereas temporally mismatched or time-averaged photometry performs worse. In this respect, the relatively compact time window of SDSS imaging may be one contributing reason why the SDSS-based sample performs somewhat better. In the application samples, this increases the likelihood that some objects may fall into regions of colour-magnitude space that are underrepresented in the training data, thereby elevating the risk of catastrophic outliers. While our S/N and quality cuts filter out the most extreme cases, they cannot eliminate this intrinsic scatter.

Third, morphology classification and the adopted photometric measurements can significantly influence the results. The point spread function (PSF) and extended (EXT) flags are assigned by survey pipelines and can exhibit instability in densely populated regions or under sub-optimal seeing conditions \citep{Hsu2014}. Blending and background-estimation errors can bias shape and flux measurements \citep[e.g. ][]{Masters2019,Melchior2018,NewmanGruen2022}. Consequently, sources such as point-like quasars blended with nearby neighbours or low-$z$ AGNs with marginally resolved hosts may be misclassified and have their colours perturbed by the adopted PSF or model photometry, pushing them into the tails of the residual distribution. The contrasts between the PSF and EXT sub-samples and between the high- and low-S/N sub-samples highlight these effects. However, the current classification methods are insufficient to identify and exclude all strongly affected objects on an individual basis. More refined diagnostics are therefore needed. In particular, LS10 residual- and fit-quality-related diagnostics, such as residual flux measurements and model-comparison statistics, may help trace source compactness, profile mismatch, and blending, and could provide useful inputs for future extensions of the model \citep{Saxena2024}. A more fundamental limitation remains for very close pairs: such systems may not be cleanly deblended, may receive biased fluxes or morphological classifications, or may not be catalogued as two separate sources. This approach therefore prioritises resolvable candidates rather than providing a complete census of close physical quasar pairs.

Finally, incomplete or non-uniform multi-band coverage, combined with our quality cuts, introduces selection effects in the training and validation sets. LS10 provides excellent wide-area photometry in the $g$, $r$, $z$, $W1$, and $W2$ bands. However, the patchy coverage in the $i$-band necessitates the exclusion of many LS10 sources with missing or noisy $i$-band measurements from the KSTD sample. Coupled with relatively stringent magnitude and error thresholds, this results in a reduction of both the effective size and colour coverage of the DESI-based samples compared to the SDSS-based samples. We also note that the adopted survey limiting magnitudes do not necessarily map one-to-one onto the effective detectability limits for quasars. In some cases, quasars may remain identifiable somewhat beyond these limits because prominent emission lines can still be detected even when the continuum is weak. This effect is not explored further here and may introduce an additional selection boundary in the present sample. Consequently, the scatter, bias, and outlier fractions reported here characterise a subset of AGNs with reasonably good photometry and complete band coverage under our adopted criteria, rather than an unbiased representation of the full quasar population \citep{NewmanGruen2022}. Extending these models will also benefit from incorporating full redshift PDFs, $p(z)$, and additional multi-band diagnostics, which help to quantify degeneracies and uncertainties that grow when band coverage is limited or incomplete \citep[e.g. ][]{Salvato2011,Salvato2019,Roster2024PICZL}.
Furthermore, it may require adjusted quality cuts and post-selection tailored to meet specific scientific objectives.

Overall, these caveats do not undermine the good performance achieved within the main redshift range and under our adopted quality cuts, but they do underscore the conditions in which AGN photo-$z$ estimates remain fragile: specifically, at high redshift, low S/N, in morphologically complex or strongly blended sources, and in regions where the spectroscopic training set is sparse or only partially reliable. Improving filter coverage, especially in the blue and near infrared wavelengths, as well as acquiring more complete and homogeneous AGN spectra in these challenging regimes, will be essential for advancing our understanding and improving the accuracy of AGN photometric redshift estimates.

\subsection{Application to the MGQPC catalogue} \label{sec5.4}
We applied our photometric redshift framework to the MGQPCs, which contained 4 112 candidate quasar pairs with Gaia DR3 astrometry and multi-wavelength photometry. For clarity, we briefly summarise the origin of this catalogue introduced in Paper~I \citep{Chen2025GaiaPairsI}: MGQPC was constructed by searching for \textit{Gaia} DR3 sources around quasars listed in the MQC within a projected distance of 100 kpc, and selecting candidate companions with astrometric properties consistent with quasars; namely, near-zero proper motions and near-zero parallaxes. The candidate sample was then further refined through additional cleaning steps and visual inspection to reject crowded stellar fields and nearby galaxies. In constructing the KSTS and KSTD training sets, we explicitly excluded all sources from the MGQPC catalogue, ensuring that the application of MGQPC serves as a strictly independent external test. Thus, in the present work, MGQPC is used only as an external application sample, whereas the supervised training is based entirely on spectroscopically confirmed quasars from SDSS, LAMOST, and DESI. For each member within the MGQPC\_A and MGQPC\_B sub-samples, we reconstructed the feature sets for KSTS and KSTD and evaluated the performance of the trained \textsc{CatBoost} and \textsc{FlexZBoost} models without any re-training or additional tuning. The resulting catalogue provided a \textsc{CatBoost} photo-$z$ estimate \texttt{z\_cat} and a \textsc{FlexZBoost} redshift PDF, $p(z)$, for each source, along with essential summary statistics, including central estimates, scatter, and credible intervals (refer to Table~\ref{tab:zphot_columns}). For the MGQPC\_A reference sub-sample, which contains reliable spectroscopic redshifts, both model configurations based on the KSTS and KSTD training samples exhibited photo-$z$ performance that is consistent with the results obtained from our comprehensive validation sets. This indicated that the models effectively transferred from the generic AGN training samples to the MGQPC context, and demonstrated their robustness and applicability across different data sets.

To prioritise candidate physical quasar pairs, we utilised our photo-$z$ predictions to assess redshift consistency within MGQPC systems. For each MGQPC, we designated the redshift of Member~A as $z_A$. When a reliable spectroscopic redshift is available, we employed that measurement; if not, we resorted to the photo-$z$ derived from the MGQPC. Member B typically lacks spectroscopic data, so we characterised it using point summaries generated by our models. Specifically, we recorded the \textsc{CatBoost} estimate $z_{{\rm Cat},B}$ and the peak redshift from the \textsc{FlexZBoost} model, denoted as $z_{{\rm Flex},B}$. This approach allowed us to systematically evaluate the redshift relationships between members of the quasar pairs, facilitating the identification of physically associated systems.

To quantify redshift consistency between the two members, we calculated the line-of-sight velocity difference, adopting Equation 10 of \citet{Hogg1999}:
\begin{equation}
    \lvert \Delta v(z_A,z_B) \rvert 
    \;=\;
    c\cdot\frac{\lvert z_A - z_B\rvert}{1 + (z_A+z_B)/2} \,,
    \label{eq:dv_def}
\end{equation} 
where $c$ is the speed of light.
Following \citet{Hennawi2010}, we adopted $\lvert \Delta v\rvert < 2\,000~\mathrm{km\,s^{-1}}$ as our pair-selection criterion. We selected systems that satisfy $\lvert \Delta v\rvert < 2\,000~\mathrm{km\,s^{-1}}$ for at least one point-estimate combination, $\lvert \Delta v(z_A, z_{{\rm Cat},B})\rvert$ or $\lvert \Delta v(z_A, z_{{\rm Flex},B})\rvert$.
The selection yielded 185 high-probability pair candidates from the MGQPC catalogue under our photo-$z$ and velocity criterion (see Table~\ref{tab:mgqpc_pairs}). We stress, however, that this candidate list is not expected to be complete for very close pairs, where blending and deblending failures may prevent such systems from being reliably measured or even catalogued as two distinct sources.

\begin{table*}
\caption{Format of the photometric redshift catalogue.}
\centering
\setlength{\tabcolsep}{5pt}
\renewcommand{\arraystretch}{1.05}
\begin{tabular}{r l l c l @{}}
\hline\hline
Column & Name & Type & Unit & Description \\
\hline
1  & GroupID     & int    & \ldots & Identifier of the candidate system \\
2  & R.A.         & double & deg    & \textit{Gaia} DR3 right ascension (ICRS J2016)  \\
3  & Decl.        & double & deg    & \textit{Gaia} DR3 declination (ICRS J2016) \\
4  & $z_{\rm Cat}$     & float  & \ldots & photometric redshift by \textsc{CatBoost}  \\
5  & $z_{\rm mean}$    & float  & \ldots & Mean of the redshift PDF \\
6  & $z_{\rm med}$     & float  & \ldots & Median of the redshift PDF \\
7  & $z_{\rm peak}$    & float  & \ldots & Peak (mode) of the redshift PDF \\
8  & $z_{\rm std}$     & float  & \ldots & Standard deviation of the redshift PDF \\
9  & num\_peaks & float  & \ldots & Number of peaks in the redshift PDF \\
10 & $z_{l68}$     & float  & \ldots & Lower bound of the 68\% credible interval \\
11 & $z_{u68}$     & float  & \ldots & Upper bound of the 68\% credible interval \\
12 & err\_u68   & float  & \ldots & Upper half-width of the 68\% interval \\
13 & err\_l68   & float  & \ldots & Lower half-width of the 68\% interval \\
\hline
\end{tabular}
\label{tab:zphot_columns}
\end{table*}

\begin{table*}
\caption{Format of the high-probability quasar pair candidates selected from the MGQPC catalogue.}
\centering
\setlength{\tabcolsep}{5pt}
\renewcommand{\arraystretch}{1.05}
\begin{tabular}{r l l c l @{}}
\hline\hline
Column & Name & Type & Unit & Description \\
\hline
1 & GroupID & int & \ldots & Identifier of the candidate system \\
2 & System Name     & string & \ldots & Pair name designated by the centre coordinates \\
3 & R.A.             & double & deg    & \textit{Gaia} DR3 right ascension (ICRS J2016)  \\
4 & Decl.            & double & deg    & \textit{Gaia} DR3 declination (ICRS J2016) \\
5 & $G$           & float  & mag    & \textit{Gaia} $G$-band magnitude \\
6 & redshift       & float  & \ldots & Adopted redshift \\
7 & Sep            & float  & arcsec & Angular separation \\
8 & $r_{p}$        & float  & kpc    & Projected separation \\
9 & $\lvert \Delta v_{r}\rvert$ & float  & km s$^{-1}$ & Line-of-sight velocity difference \\
\hline
\end{tabular}
\begin{minipage}{0.95\linewidth}
\footnotesize
\raggedright
\vspace{3pt}
\textbf{Notes.} Rows with the same GroupID belong to the same candidate system. Each system occupies two rows, corresponding to Member A and Member B.
\end{minipage}
\label{tab:mgqpc_pairs}
\end{table*}

As an external check, we cross-matched the 185 high-probability MGQPC systems with the currently available spectroscopically identified MGQPC systems from \citet{Deng2025GaiaPairsIII} and \citet{Jing2026}. The former reported 18 spectroscopically confirmed quasar pairs and 142 projected quasars based on DESI DR1, while the latter independently confirmed 49 quasar pairs contained in the MGQPC catalogue. Together, these data provide 67 spectroscopically confirmed quasar pairs and 142 projected quasars for external comparison. Under our fiducial cut of $|\Delta v_{\rm phot}| < 2\,000~\mathrm{km~s^{-1}}$, cross-matching the 185 candidates with the external spectroscopic comparison sample yielded 20 confirmed quasar pairs and 8 projected quasars (Table~\ref{tab:mgqpc_confirmed_full}); representative examples are shown in Fig.~\ref{fig:mgqpc_confirmed}. Relaxing the threshold to $|\Delta v_{\rm phot}| < 3\,000~\mathrm{km~s^{-1}}$ increased the candidate list to 288 and yielded 28 confirmed pairs and 8 projected quasars. Thus, relaxing the velocity threshold recovered more quasar pairs, but may also admit additional projected contaminants. This behaviour is expected because these velocity windows correspond to a very tight redshift tolerance (of order $|z_1-z_2| \approx 0.01$--$0.03$ at $z \approx 1$--$2$), comparable to or smaller than the intrinsic scatter of AGN photo-$z$ estimates, especially for close and potentially blended systems. Overall, the comparison with the external spectroscopic sample provides additional validation of our prioritisation scheme and supports combining \textsc{CatBoost} point estimates with \textsc{FlexZBoost} redshift PDFs to prioritise quasar pair candidates for follow-up observations.

\section{Conclusions} \label{sec6}
We constructed two well-defined spectroscopic samples (KSTS and KSTD) utilising data from SDSS DR18, LAMOST DR12, and DESI DR1, complemented by homogeneous multi-band photometry from SDSS, DESI-LS, \textit{WISE}, and \textit{Gaia}. Based on these samples, we utilised SHAP-guided feature selection to identify a compact and physically interpretable set of input features, and then trained the \textsc{CatBoost} algorithm to generate point-estimate photo-$z$ values and employed \textsc{FlexZBoost} to produce comprehensive redshift PDFs. The resulting models demonstrate competitive accuracy across a broad redshift range, exhibiting minimal residual biases and well-behaved PDF diagnostics. This indicates that they provide both robust point estimates and well-calibrated redshift probability distributions for AGNs in wide-area imaging surveys. Control tests indicate that the better performance of the SDSS-based sample should not be interpreted purely as a model effect, but as reflecting both the added $u$-band information and survey-dependent differences in the underlying photometric data products, with the $u$-band effect likely being the main factor and LS10 bright-source systematics contributing at a secondary level. However, as is discussed in Sect.~5.3, the overall performance is constrained in several contexts: high-redshift regimes, low-S/N, morphologically complex, or blended sources, and areas where the spectroscopic training set is sparse or affected by misidentified redshifts. These limitations represent the dominant sources of uncertainty within our framework and should be carefully considered when interpreting photo-$z$ statistics, particularly for extreme objects.

We subsequently applied this framework to the MGQPC as a comprehensive external test, ensuring that all MGQPC sources were excluded from the training samples. For each component within the MGQPC, we computed a photo-$z$ using \textsc{CatBoost}, generated a redshift PDF with \textsc{FlexZBoost}, and derived summary statistics, including the peak redshift of the PDF. We designated the MGQPC redshift for the primary component and utilised our photo-$z$ estimate for the secondary component to compute the line-of-sight velocity difference, selecting candidate pairs with $\lvert \Delta v \rvert < 2\,000~\mathrm{km\,s^{-1}}$. This filtering process yielded 185 MGQPC systems that are photometrically consistent with being quasar pairs. We then investigated which of these candidates have associated spectroscopic data. Among them, 20 are matched to spectroscopically confirmed quasar pairs and 8 are matched to projected quasars in the current external spectroscopic comparison sample. These matches provide an external validation of the effectiveness of our photo-$z$ based filtering strategy.

Collectively, our findings indicate that integrating gradient-boosted decision trees for point estimates with flexible CDE for comprehensive redshift PDFs establishes a practical framework for quasar photo-$z$ determination and quasar pair searches within extensive photometric catalogues. The MGQPC photometric redshift catalogue constructed in this work, along with the associated high-probability pair candidates, offers a valuable foundation for planning future spectroscopic confirmation of dual and projected quasars. In recent years, and during the course of this work, spectroscopic follow-up observations of several quasar pair candidates have also been carried out. The results of these observations may help verify or improve our quasar pair selection. The relevant follow-up results will be presented in a forthcoming paper. Looking ahead, enhancing the training samples to include fainter and more diverse AGN populations, incorporating additional wavelength coverage and time-domain information, and further refining quasar pair selection strategies could significantly improve the efficiency and reliability of quasar pair searches in forthcoming surveys such as LSST, \textit{Euclid}, and those of the Chinese Space Station Telescope (\textit{CSST}).

\section*{Data Availability}
Tables~\ref{tab:zphot_columns} and~\ref{tab:mgqpc_pairs} are only available in electronic form at the CDS via anonymous ftp to \url{cdsarc.u-strasbg.fr} (130.79.128.5) or via \url{http://cdsweb.u-strasbg.fr/cgi-bin/qcat?J/A+A/}. The original MGQPC catalogue can be provided by the corresponding author upon reasonable request.

\begin{acknowledgements}
The authors thank the anonymous referee for the valuable comments that improved the quality and clarity of the manuscript. This work has been supported by the National Key R\&D Program of China (2025YFA1614101, 2021YFA0718500) and by the Chinese National Natural Science Foundation grant No. 12333001. J.-Q.~Xia acknowledges support from the China Manned Space Program (grant Nos.~CMS-CSST-2025-A01 and CMS-CSST-2025-A04).

This work has made use of data from the ESA \textit{Gaia} mission (\url{https://www.cosmos.esa.int/gaia}), processed by the \textit{Gaia} Data
Processing and Analysis Consortium (DPAC; \url{https://www.cosmos.esa.int/web/gaia/dpac/consortium}), with funding provided by the institutions participating in the \textit{Gaia} Multilateral Agreement.

Funding for the Sloan Digital Sky Survey V has been provided by the Alfred P.\ Sloan Foundation, the Heising Simons Foundation, the National Science Foundation, and the Participating Institutions. SDSS acknowledges support from the Centre for High Performance Computing at the University of Utah. The SDSS website is \url{www.sdss.org}.

LAMOST (the Guoshoujing Telescope) is a National Major Scientific Project built by the Chinese Academy of Sciences, funded by the National Development and Reform Commission, and operated by the National Astronomical Observatories of the Chinese Academy of Sciences.

This research used data obtained with the Dark Energy Spectroscopic Instrument (DESI).
DESI is managed by Lawrence Berkeley National Laboratory and supported by the U.S.\ Department of Energy Office of Science, the U.S.\ National Science Foundation, and DESI member institutions (\url{www.desi.lbl.gov/collaborating-institutions}). We acknowledge the significance of I'oligam Du'ag (Kitt Peak) to the Tohono O'odham Nation.

The DESI Legacy Imaging Surveys comprise DECaLS, BASS, and MzLS. Data processing was supported by NOIRLab and Lawrence Berkeley National Laboratory. The Legacy Surveys were supported by the U.S.\ Department of Energy, NERSC, the National Science Foundation, and supporting institutions in the United States and China.

We acknowledge the use of public data from the following facilities: \textit{Gaia}, SDSS, LAMOST, DESI, DESI Legacy Survey. This research made use of the following open-source software:
Astropy \citep{AstropyCollaboration2013,AstropyCollaboration2018,AstropyCollaboration2022},
Matplotlib \citep{Hunter2007Matplotlib},
NumPy \citep{Walt2011NumPy,Harris2020NumPy},
pandas \citep{McKinney2010Pandas,pandas2022},
and TOPCAT \citep{Taylor2005TOPCAT}.

\end{acknowledgements}

\bibliographystyle{aa}
\bibliography{aa58953-26}

\clearpage
\onecolumn
\begin{appendix}

\begingroup
\setlength{\parskip}{0pt}
\setlength{\parindent}{0pt}
\setlength{\abovecaptionskip}{1pt}
\setlength{\belowcaptionskip}{0pt}

\section{Photometric features}
\label{app:features}

Candidate photometric features were divided into three categories: model magnitudes in all available bands, colours from adjacent and non-adjacent bands, and same-aperture adjacent-band colours.

\vspace{0.20em}

{\footnotesize
\noindent\textit{KSTS (23 features):}
$u-g$, $W1-G$, $W1$, $W2-G$, $g-r$, $z-W2$, $u-r$, $W1-G_{\rm RP}$, $r-i$,
$W1-W2$, $G_{\rm BP}-G$, type, $r-z$, $z-W1$, $i-W1$, $i-z$, $u-G$, $g-i$,
$u$, $W2$, $G-G_{\rm RP}$, $r-W2$, and $r$.

\vspace{0.20em}

\noindent\textit{KSTD (53 features):}
$z-W2$, $r-z$, $W1-W2\_ap1$, $G_{\rm BP}-G$, $g-r$, $z-W1$, $g-G$, $W1$,
$W1-W2$, $r-G$, $i-z$, $W1-G$, $r-i$, $z-W2\_ap1$, $W2\_ap1$, $W1\_ap1$,
$z-G_{\rm RP}$, $G_{\rm BP}$, $g$, $G-G_{\rm RP}$, $W1-G_{\rm RP}$, $i-W1$,
$W2-G$, $z-G$, $W1-W2\_ap2$, $r-z\_ap6$, $W1-W2\_ap5$, $z-W1\_ap1$,
$r-z\_ap5$, $r-i\_ap5$, $g-r\_ap5$, $z-W2\_ap4$, $g-r\_ap6$, $g-r\_ap8$,
$g\_ap8$, $r-W2\_ap1$, $z\_ap4$, $W2$, $r-W2$, $g-W2\_ap1$, $r$,
$z-W2\_ap2$, $g-G_{\rm BP}$, type, $z-W2\_ap3$, $G_{\rm RP}$, $g-W1$,
$g-z$, $r-W2\_ap4$, $W1-G_{\rm BP}$, $g-W2$, $r-W2\_ap2$, and $g-W2\_ap4$.
}

\vspace{-0.80em}

\section{Externally matched spectroscopic systems}
\label{app:external_match}

\vspace{-0.50em}

\centering
\includegraphics[width=0.499\textwidth]{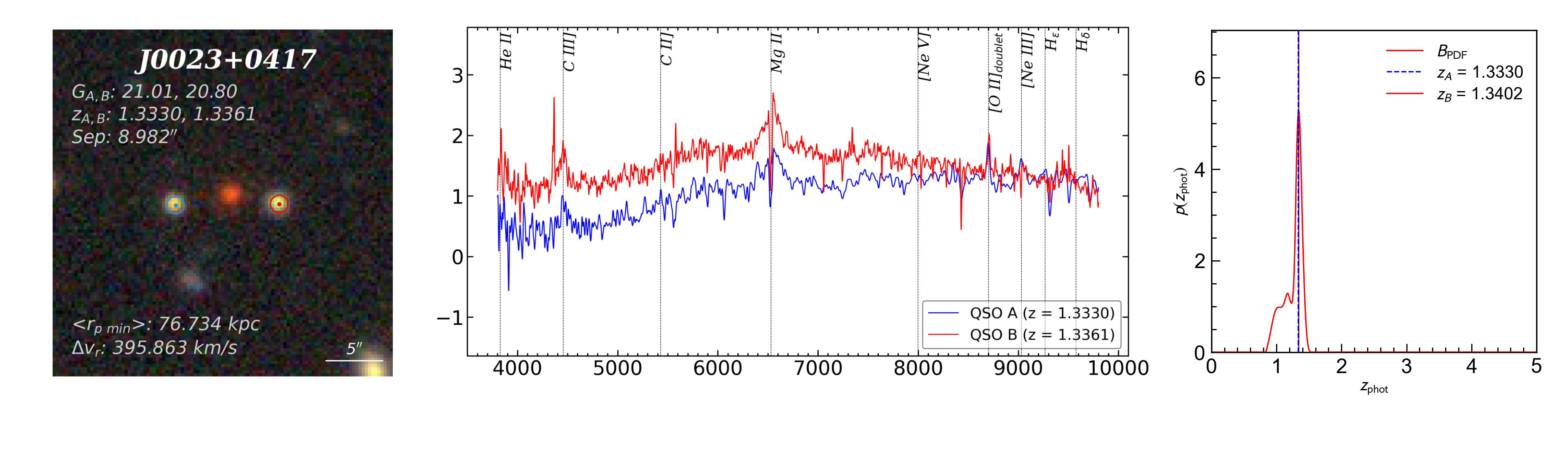}\hspace{-0.2em}
\includegraphics[width=0.499\textwidth]{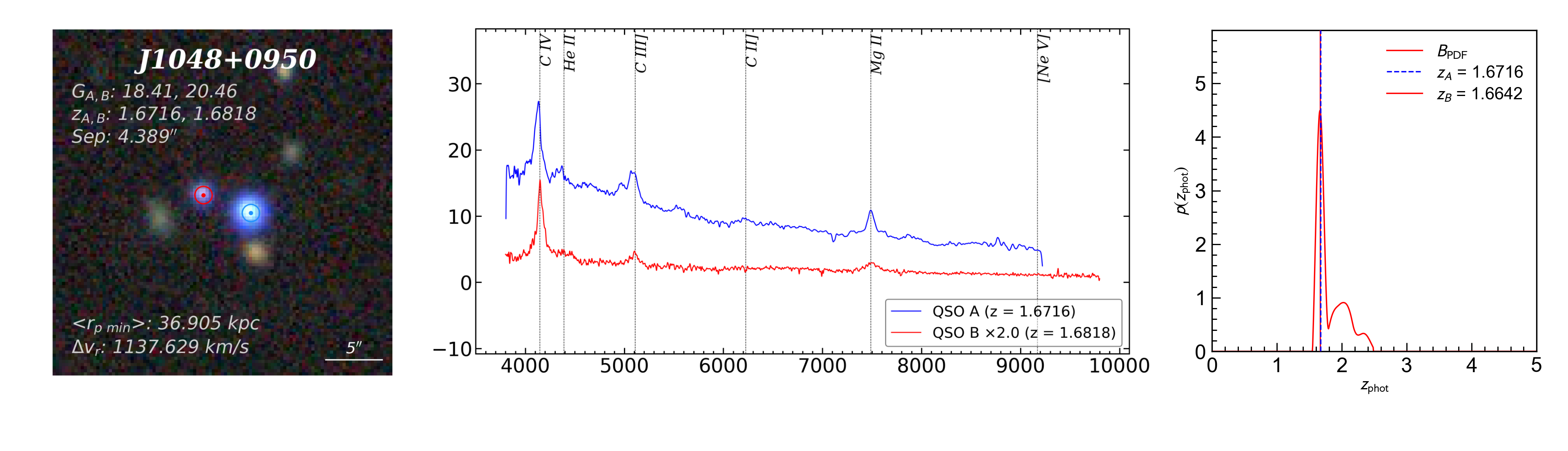}

\vspace{-1.20em}
\includegraphics[width=0.499\textwidth]{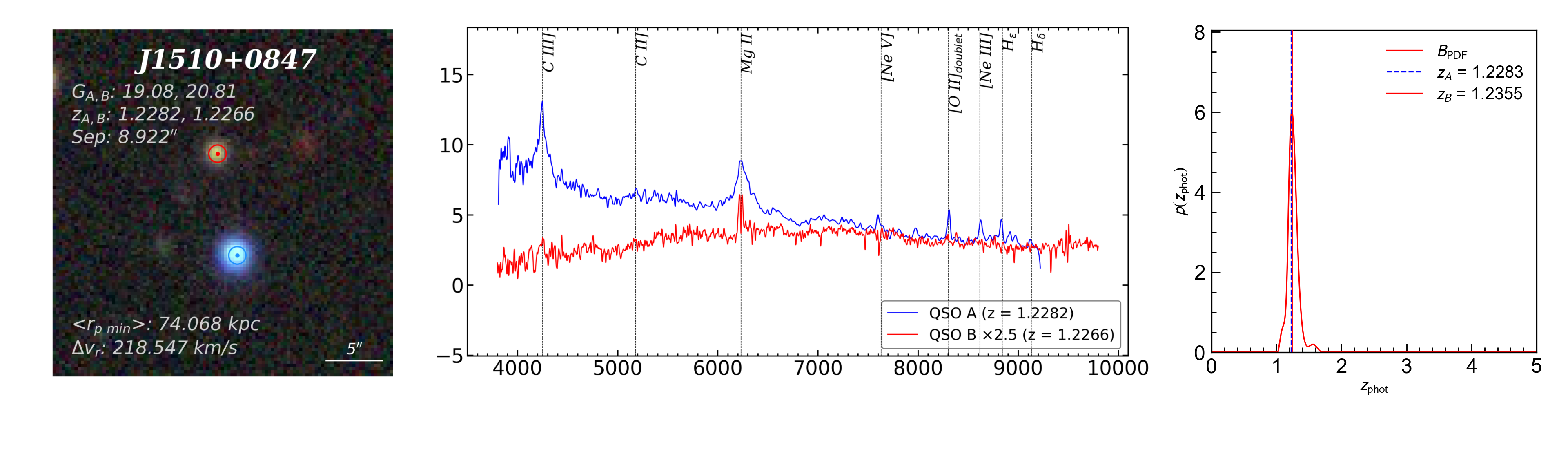}\hspace{-0.2em}
\includegraphics[width=0.499\textwidth]{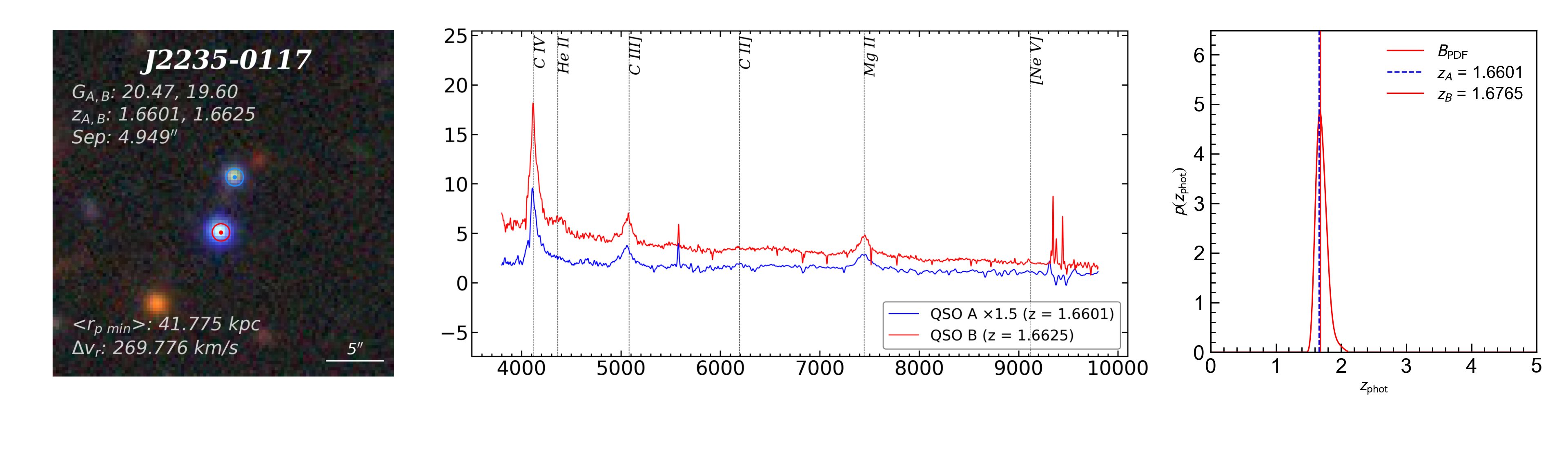}

\vspace{-1.20em}
\includegraphics[width=0.499\textwidth]{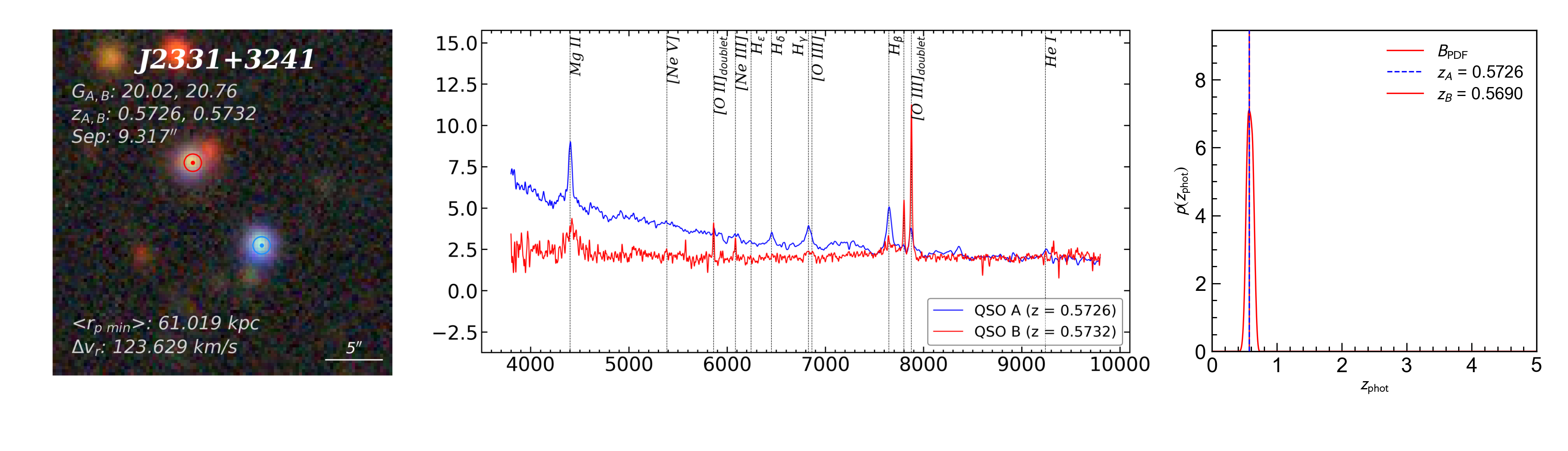}\hspace{-0.2em}
\includegraphics[width=0.499\textwidth]{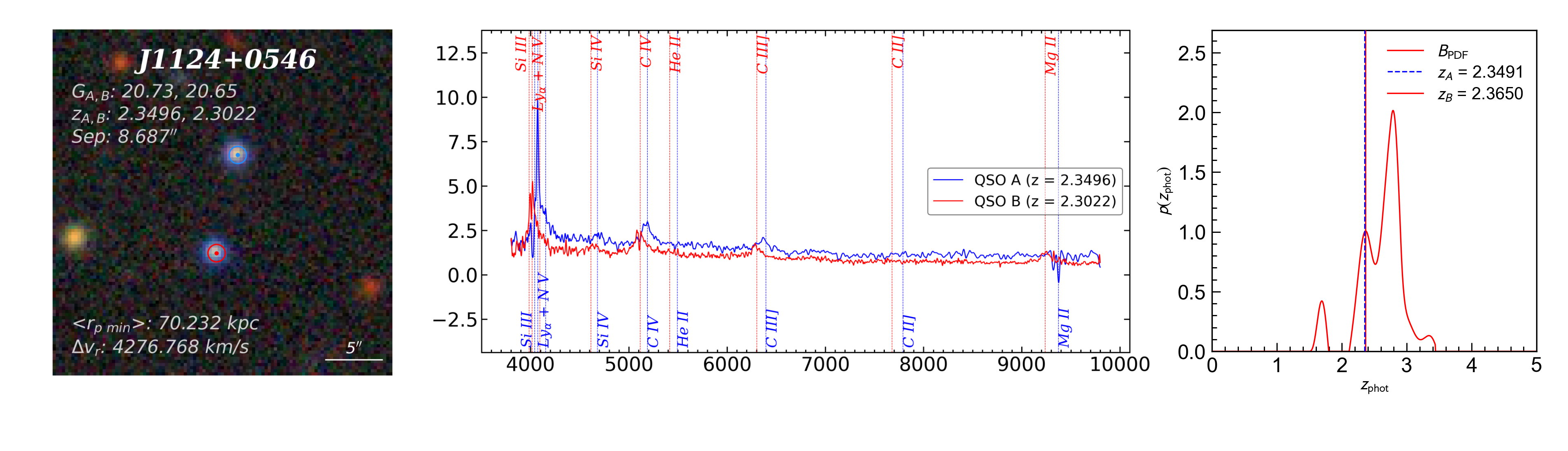}

\vspace{-1.05em}

\renewcommand{\thefigure}{B.\arabic{figure}}
\setcounter{figure}{0}
\refstepcounter{figure}
\label{fig:mgqpc_confirmed}
\raggedright
{\footnotesize\textbf{Fig.~B.1.} Representative spectroscopically identified systems within our high-probability MGQPC sample. Each row shows the image cutout, the spectra of Member A and Member B, and the corresponding photo-$z$ information, with available spec-$z$ annotations.}

\vspace{0.50em}

\renewcommand{\thetable}{B.\arabic{table}}
\setcounter{table}{0}

\captionof{table}{Spectroscopically identified systems within our high-probability MGQPC candidate sample with externally adopted $z_{\rm pB}$.}
\label{tab:mgqpc_confirmed_full}

\vspace{-0.20em}

\centering
{\small
\setlength{\tabcolsep}{5pt}
\renewcommand{\arraystretch}{1.05}
\begin{tabular}{l c c c c c c}
\hline\hline
System Name &
$z_A$ &
$z_B$ &
$z_{\rm pB}$ &
Sep (arcsec) &
$r_p$ (kpc) &
$\lvert\Delta v_r\rvert$ (km s$^{-1}$)\\
\hline
J002337.865$+$041700.483 & 1.333 & 1.336 & 1.340 & 9.136 & 78.775 & 395.590 \\
J104846.347$+$095012.218 & 1.672 & 1.682 & 1.664 & 4.358 & 37.849 & 1137.629 \\
J145758.778$+$251648.214 & 1.374 & 1.386 & 1.387 & 5.674 & 49.055 & 1430.754 \\
J151050.693$+$084718.324 & 1.228 & 1.227 & 1.236 & 8.903 & 76.064 & 218.547 \\
J233100.605$+$324142.896 & 0.573 & 0.573 & 0.569 & 9.331 & 62.873 & 123.629 \\
J003912.916$+$053407.666 & 2.073 & 2.083 & 2.061 & 4.953 & 42.290 & 1000.177 \\
J073630.828$+$263729.719 & 1.635 & 1.634 & 1.632 & 9.532 & 82.818 & 127.000 \\
J092217.534$-$011752.748 & 1.689 & 1.682 & 1.666 & 6.030 & 52.360 & 750.045 \\
J133741.828$+$030029.891 & 1.563 & 1.566 & 1.575 & 10.369 & 90.101 & 284.278 \\
J144003.256$+$390834.669 & 1.277 & 1.279 & 1.272 & 11.234 & 96.445 & 229.961 \\
J153701.065$+$322326.785 & 1.828 & 1.827 & 1.831 & 0.970 & 8.388 & 89.039 \\
J014757.104$+$184342.137 & 1.587 & 1.583 & 1.602 & 8.742 & 75.971 & 482.975 \\
J075354.493$+$181448.115 & 2.382 & 2.372 & 2.392 & 5.115 & 42.720 & 849.016 \\
J082017.146$+$265214.835 & 2.009 & 2.016 & 1.996 & 8.366 & 71.716 & 722.937 \\
J090824.299$+$033926.887 & 1.270 & 1.279 & 1.255 & 7.658 & 65.726 & 1238.480 \\
J155057.281$+$022146.524 & 2.398 & 2.397 & 2.387 & 2.027 & 16.899 & 62.705 \\
J223550.970$-$011739.994 & 1.660 & 1.663 & 1.677 & 4.932 & 42.842 & 269.776 \\
J074013.437$+$292647.058 & 0.977 & 0.980 & 0.968 & 2.653 & 21.730 & 408.630 \\
J090449.516$-$012210.285 & 0.835 & 0.832 & 0.823 & 3.298 & 25.823 & 461.071 \\
J131012.604$+$042106.814 & 2.119 & 2.108 & 2.123 & 9.418 & 80.224 & 1010.183 \\
\hline
J112418.823$+$054654.239 & 2.350 & 2.302 & 2.365 & 8.593 & 72.067 & 4276.767 \\
J114319.420$+$031715.421 & 0.696 & 0.535 & 0.690 & 8.893 & 61.937 & 29978.782 \\
J125723.575$+$005551.486 & 1.325 & 1.343 & 1.321 & 11.060 & 95.360 & 2229.681 \\
J135038.967$-$030115.128 & 2.813 & 2.770 & 2.826 & 3.319 & 26.703 & 3345.532 \\
J145453.924$+$040929.208 & 2.879 & 2.146 & 2.891 & 11.441 & 94.488 & 62534.974 \\
J160843.942$+$055615.788 & 1.868 & 2.005 & 1.857 & 6.836 & 58.837 & 13966.705 \\
J162113.227$+$230407.847 & 1.505 & 1.533 & 1.494 & 9.382 & 81.485 & 3315.037 \\
J231322.351$+$070511.036 & 1.140 & 1.862 & 1.867 & 3.791 & 32.914 & 86544.680 \\
\hline
\end{tabular}
}
\begin{minipage}{0.95\linewidth}
\footnotesize
\raggedright
\vspace{2pt}
\textbf{Notes.} $z_{\rm pB}$ is the externally adopted redshift for Member~B. The first 20 rows list spectroscopically confirmed quasar pairs, and the last 8 rows list projected quasars.
\end{minipage}

\endgroup
\end{appendix}

\end{document}